\documentclass[conference,onecolumn]{IEEEtran}

\usepackage{swami}

\newtheorem{defn}{\bf Definition}
\newtheorem{lemma}{Lemma}
\newtheorem{thm}{Theorem}

\newtheorem{rem}{Remark}

\newcommand{\G} {{\mathcal{G}}}

\renewcommand{\L} {{\mathcal{L}}}

\renewcommand{\S} {{\mathcal{S}}}

\newcommand{\E} {{\mathcal{E}}}
\newcommand{\C} {{\mathcal{C}}}

\newcommand{\T} {{\mathcal{T}}}
\newcommand{\M} {{\mathcal{M}}}

\newcommand{\commentswami}[1] {}

\author
{
\IEEEauthorblockN{
Swaminathan Sankararaman\IEEEauthorrefmark{1},
Alon Efrat\IEEEauthorrefmark{1},
Srinivasan Ramasubramanian\IEEEauthorrefmark{2} and
Javad Taheri\IEEEauthorrefmark{1}
}
\IEEEauthorblockA{ 
\IEEEauthorrefmark{1}Department of Computer Science \\
\{swami,alon,taheri\}@cs.arizona.edu
}
\IEEEauthorblockA{
\IEEEauthorrefmark{2}Department of Electrical and Computer Engineering \\
srini@ece.arizona.edu
}
\it{University of Arizona, Tucson}
, AZ 85721 
}

\title{Scheduling Sensors for Guaranteed Sparse Coverage}

\begin{document}
\maketitle

\begin{abstract}
Sensor networks are particularly applicable to the tracking of objects in motion. For such applications, it may not necessary that the whole region be covered by sensors as long as the uncovered region is not too large. This notion has been formalized by Balasubramanian et.al. \cite{ravi-coverage} as the problem of $\kappa$-weak coverage. This model of coverage provides guarantees about the regions in which the objects may move undetected. In this paper, we analyse the theoretical aspects of the problem and provide guarantees about the lifetime achievable. We introduce a number of practical algorithms and analyse their significance. The main contribution is a novel linear programming based algorithm which provides near-optimal lifetime. Through extensive experimentation, we analyse the performance of these algorithms based on several parameters defined.
\end{abstract}

\section{Introduction}
\label{sec:intro}

Wireless sensor networks are being increasingly used to due to its vast spectrum of applications including surveillance, security, healthcare, asset tracking and environmental monitoring. Such a network consists of large numbers of nodes which are placed in a region in order to gather and transmit information about activities in that region. Each node is a low power device consisting of sensors and communication units with limited processing power. In general, since sensor networks deployment and replacement of nodes in a sensor network is an expensive process, it is of great interest to conserve energy resources in the network and extend the lifetime of the network.

The coverage problem in sensor networks (see \cite{ghosh-covsurvey,Ilyas-handbooksensornw}) has been extensively studied in literature. The worst-case coverage problem may be defined in terms of two related problems: (i) \emph{The Maximal Breach problem}, discussed in \cite{meguer-cov} and (ii) \emph{The Minimal Exposure problem}, discussed in \cite{megerian-minexposurecov}. The Maximal Breach problem defines it in terms of the distance of an object to be tracked from any sensor and the Minimal Exposure problem defines it by using the sensitivity of a point by a sensor and finding the integral of the sensitivities of the points on the path of a moving object to be tracked. In both cases, paths where these parameters have lowest values are used to characterize the coverage of a region.

In many regions, sensor nodes are deployed in large numbers randomly and this generates a lot of redundancy in the network. In order to maximize the coverage time, sensors may be turned on and off so that coverage is still maintained. Various techniques have been proposed for maximizing the lifetime by allowing certain nodes to sleep (see \cite{fullcov-maxlife, xing-covsensornw, wu-covtime}). In \cite{ravi-coverage}, the authors investigate the coverage time characteristics of sensor networks and develop algorithms to provide optimal coverage time when sensors are allowed to be turned on and off.

With regard to sparse coverage, most of the work has used the notion of fractional coverage, i.e., only a fraction of the region is allowed to be uncovered. The majority of such research involves trying to make the coverage holes occupy as little area as possible (see \cite{wan-kcov, zhang-alphalifetime}). 

In general, it may be advantageous to model sparse coverage using the diameter of coverage holes instead of area. For example, in surveillance applications, it may be necessary to restrict the motion of a target under surveillance. The target may be restricted to only move a certain distance before it is detected. In order to solve this problem, the notion of \textbf{\emph{$\kappa$-weak coverage}} was introduced by Balasubramanian et. al.\cite{ravi-coverage}. Recently, it was independently re-introduced by Balister et.al. \cite{sinha-trapcoverageinfocom09} as \textbf{\emph{trap coverage}} and was investigated for a random deployment of sensors. To the best of our knowledge, \cite{ravi-coverage} was the first work to introduce this model of coverage and provided a greedy algorithm for achieving $\kappa$-weak coverage. In $\kappa$-weak coverage, the parameter $\kappa$ denotes the restriction on the diameter of the holes.

There are two interesting points in the definition of $\kappa$-weak coverage, when compared to full coverage, on top of the much longer lifetime.

\begin{enumerate}
\item It is easy to compute, and as our experiments indicate, much easier to give a PTAS (polynomial time approximation scheme) to within factor of $1+\eps$ for the coverage holes.
\item It is easier to compute in a distributed manner.
\end{enumerate}

\textbf{Our Contributions:} In this paper, we study several practical algorithms for $\kappa$-weak coverage, with theoretical guarantees. We introduce several algorithms to find $\kappa$-weak coverage including a novel algorithmic correlation between network flow algorithms and $\kappa$-weak coverage. We provide a Linear Programming algorithm based on this correlation.

In all settings, it is important to distinguish between the preemptive and non-preemptive cases. In the former, we can turn sensors on and off (activate and deactivate) as long as their total active time is at most their original battery life and in the latter, once a sensor is activated, it remains active till its battery is depleted. In this paper, we assume that their original battery is, for convenience, uniform, but in  general is does not have to be the case.  We will also normalize such that if a sensor is working for time $\delta$ than its battery loses $\delta\%$. We also assume that the range of communication is twice the range of sensing. So two sensors can communicate if and only if their sensing regions overlap. 

The rest of the paper is organized as follows. Section \ref{sec:form} deals with the formalization of the problem and the objectives of this paper. Section \ref{sec:theo} goes over the techniques employed and the lifetime guarantees achievable. Section \ref{sec:findcover} talks about the algorithms to find $\kappa$-weak covers together with their scheduling in detail. Section \ref{sec:results} discusses the simulation results and in Section \ref{sec:conc}, we provide the conclusion.
\section{Problem Formulation}
\label{sec:form}

The network model consists of a set of sensors $\S$ which are placed
in a region of interest $R$ known as the \textbf{\emph{Sensor
  Field}}.  The sensing range of each sensor $i$ located at $(x,y)$, is
a disk of radius $1$ centered at its location and is denoted by
$D_{i}$. Initially, every sensor has the same battery life and this is
assumed to be $1$ unit of time. A point $p$ is said to be covered if
$p\in D_{i}$ for some sensor $i$. From now on, when we say a sensor $i$
covers a region, it implies that the region is contained in $D_{i}$. We restrict ourselves to square environments of dimensions $s\times s$.

\begin{defn}
 Let $\kappa > 1$ be a fixed parameter. We say that a pair of points
 $(p,q)$ is a \textbf{\emph{$\kappa$-pair}} if the distance between them is
 $\kappa$. A curve $\mu$ is a \textbf{\emph{$\kappa$-curve}} is its endpoints form a
 $\kappa$-pair. A set of sensors $C\subseteq \S$ is a \textbf{\emph{$\kappa$-weak
      cover}} if every $\kappa$-curve that is contained in $R$
  intersects $\displaystyle \bigcup_{s\in C} D_{s}$.
\end{defn}

The problem under consideration is to obtain a collection
$\C=\{C_{1},C_{2},\dots,C_{k}\}$ of subsets of $\S$ together with a schedule
$\Delta=\{\delta_{1},\delta_{2},\dots,\delta_{k}\}$ where each $C_{i}$ is a
$\kappa$-weak cover and $\displaystyle \L=\sum_{i=1}^{k} \delta_{i}$ is
maximized. The parameter $\L$ is termed as the \textbf{\emph{Lifetime}}
of the system. 

We are also interested in finding a PTAS for the above
problem. In this case, we are interested in finding $\C$ and $\Delta$
such that each $C_{i}\in \C$ is a $((1+\epsilon)\kappa)$-weak cover.

In addition, we define the following types of schedules.

\begin{defn}
 A schedule $\Delta$ is said to be (i) \textbf{\emph{Non-Preemptive}} if
 for any $\delta_{i}\in \Delta$, $\delta_{i}=1$, i.e., each cover
 $C_{i}$ is kept ON till the batteries are used up, (ii)
 \textbf{\emph{Uniform}} if for any $\delta_{i},\delta_{j}\in \Delta$, 
 $\delta_{i}=\delta_{j}$, i.e., all covers are kept ON for the same
 amount of time and (iii) \textbf{\emph{Non-Uniform}} if for some
 $\delta_{i},\delta_{j}\in \Delta$, $\delta_{i}\neq \delta_{j}$,
 i.e., all covers need not be kept ON for the same amount of time.
\end{defn}

\section{Overview and Lifetime Guarantees}
\label{sec:theo}

We first give an upper bound for the optimal lifetime of the
system. Then, we proceed to outline the motivation for the Grid-Based
Techniques and prove some useful properties of these techniques which
provides a guarantee of the lifetime of the system under these techniques.

\begin{defn}
  The following definitions are with respect to a set of sensors
  $\S$. The \textbf{\emph{depth of a point $p$}}, denoted by $d_{p}$
  is defined as the number of sensors in $\S$ which cover $p$. The
  \textbf{\emph{depth of a region $R$}} denoted by $d_{R}$ is 
  defined to be the maximum depth of any point $p$ in $R$.
\end{defn}

Let $\L^{*}$ denote the maximum lifetime of the system under optimal
scheduling. The following lemma gives an upper bound for $\L^{*}$.

\begin{thm}
\label{thm:optL}
  $\L^{*} = O(\kappa\cdot d_{R})$.
\end{thm}

\begin{IEEEproof}
Let $\C^{*}=\{C_{1}^{*},C_{2}^{*},\dots,C_{k^{*}}^{*}\}$ be the covers
and $\Delta^{*}=\{\delta_{1}^{*},\delta_{2}^{*},\dots,\delta_{k^{*}}^{*}\}$
be the schedule which provide the optimal lifetime $L^{*}$, so that
$\displaystyle \sum_{i=1}^{k^{*}} \delta_{i}^{*} = L^{*}$. Pick an
arbitrary line $l$ and assume that $f$ is the interval between the two
most remote sensors along $l$. In each time interval $\delta_{i}^{*}$,
let $s_{a},s_{b}\in C_{i}^{*}$ be two sensors along $l$ such that
there is no sensor which covers the portion of the line between
$s_{a}$ and $s_{b}$. The distance between $s_{a}$ and $s_{b}$ is
atmost $\kappa$, otherwise $C_{i}^{*}$ is not $\kappa$-weak.

Hence, the minimum number of sensors in $C_{i}^{*}$ which intersect
$l$ is $\frac{f}{\kappa}$ and the total battery life used up for
$\delta_{i}^{*}$ is $\delta_{i}^{*}\cdot \frac{f}{\kappa}$. Now, the
the total battery at the start of the schedule for the set of sensors
intersecting $l$ is atmost $f\cdot d_{R}$. Therefore,

\begin{IEEEeqnarray}{rCl}
\sum_{i=1}^{k^{*}} \delta_{i}^{*}\cdot \frac{f}{\kappa} & \leq & f\cdot
d_{R} \nonumber \\
\sum_{i=1}^{k^{*}} \delta_{i}^{*} & \leq & \kappa\cdot
  d_{R} \nonumber \\
L^{*} & = & O(\kappa\cdot d_{R}) \label{eqn:optbound}
\end{IEEEeqnarray}
\end{IEEEproof}

In this paper, we formulate two techniques in order to find $\C$ and $\Delta$: (i) Grid-Based and (ii) Random Seeds.

\subsection{Grid-Based}
\label{sec:gridoverview}

Consider a tiling $T$ of $R$ in which each component of the tiling has diameter of at most $\kappa$. Now, it is clear that any subset of sensors $C$ that completely covers the boundaries of $T$ is a $\kappa$-weak cover for $R$. The rest of this section investigates the lifetime of the system by generating $C$ from such tilings. 

\begin{defn}
 The \textbf{\emph{Total Edge Length}} of a tiling $\T=\{T_{1},T_{2},\dots,T_{n}\}$, denoted by $TEL_{\T}$ is defined as the sum of the lengths of the boundaries of the tiles in $\T$. More formally, if $l_{T}$ denotes the length of the boundary of $T$, then,
\[
\displaystyle
TEL_{\T} = \sum_{T\in \T} l_{T}
\]
\end{defn}

Intuitively, for any tiling $\T$, if $\kappa\gg 1$ then in any cover $C$ for $\T$, $|C|$ is directly proportional to $TEL_{\T}$. 

We investigate the properties of two tilings: (i) the square tiling and (ii) the hexagonal tiling. From now, whenever we refer to a grid, it implies that we are referring to the edges of a tiling.

\begin{defn}
  A \textbf{\emph{Square Grid}} for a region $R$ of dimensions $L\times L$ is defined to be a set of vertical and horizontal line segments $\{l_{1},l_{2},\dots,l_{m}\}$ such that each $l_{i}$ is of length $L$ and each square of the implied tiling is of dimensions $\frac{\kappa}{\sqrt{2}}\times \frac{\kappa}{\sqrt{2}}$ (see Figure \ref{fig:squareshift} for examples of such grids).
\end{defn}

\begin{defn}
  A \textbf{\emph{Hexagonal Grid}} for a region $R$ of dimensions $L\times L$ is defined to be a set of line segments $\{l_{1},l_{2},\dots,l_{m}\}$ such that each $l_{i}$ is an edge of a hexagonal tiling superimposed on $R$ where each hexagon is of diameter $\kappa$ (see Figure \ref{fig:hexshift} for examples of such grids). Hence, the length of each $l_{i}$ is $\frac{\kappa}{2}$.
\end{defn}

\begin{lemma}
\label{lem:TELsq}
 The TEL of the square tiling is $4\sqrt{2}\frac{L^{2}}{\kappa}$.
\end{lemma}

\begin{IEEEproof}
The size of each square is $\frac{\kappa}{\sqrt{2}}$. Hence, the number of squares inside an $L\times L$ region is $\frac{2L^{2}}{\kappa^{2}}$. Since the perimeter of each square is $4\frac{\kappa}{\sqrt{2}}$, we have
\[
TEL_{Square Grid} = \frac{2L^{2}}{\kappa^{2}}\cdot 4\frac{\kappa}{\sqrt{2}}
\]
since each edge is counted twice. Hence, $TEL=4\sqrt{2}\frac{L^{2}}{\kappa}$.

\end{IEEEproof}

\begin{lemma}
\label{lem:TELhex}
 The TEL of the hexagonal tiling is $\frac{8}{\sqrt{3}}\frac{L^{2}}{\kappa}$.
\end{lemma}

\begin{IEEEproof}
The side of each hexagon is $\frac{\kappa}{2}$ since its diameter is $\kappa$ and its area is $\frac{3\sqrt{3}}{8}\cdot\kappa^{2}$. Hence, the number of hexagons inside an $L\times L$ region is $\frac{8}{3\sqrt{3}}\cdot\frac{L^{2}}{\kappa^{2}}$. Since the perimeter of each hexagon is $3\kappa$, we have
\[
TEL_{Hexagonal Grid} = \frac{8}{3\sqrt{3}}\cdot \frac{L^{2}}{\kappa^{2}}\cdot 3\kappa
\]
since each edge is counted twice. Hence, $TEL_{Hexagonal Grid}=\frac{8}{\sqrt{3}}\frac{L^{2}}{\kappa}$.
\end{IEEEproof}

As input to the algorithm, we are given a collection of such grids $\G=\{G_{1},G_{2},\dots,G_{l}\}$ and as output, we find a set of covers $\C$ such that each subset $C_{i}\in \C$ is a cover for one of the grids $G_{j}\in \G$. The algorithm iterates over all the grids repeatedly until no cover is found for all the grids in $\G$. The steps of the algorithm are outlined in Figure \ref{fig:gridoutline}. In essence, there are two parts to this algorithm: (i) Obtaining the collection $\G$ and (ii) Finding a cover for a grid $G$. We describe (i) here and (ii) is described in Section \ref{sec:findcover}. Further, we also describe how to find $\Delta$ in Section \ref{sec:findcover}.

\begin{rem}
\label{rem:lifetimecriteria}
We may infer that the two criteria for obtaining maximum lifetime from a collection of grids $\G$ are (i) $|\G|$ and (ii) The number of sensors common between any two grids $G_{i},G_{j}\in \G$.  If the second criteria is minimized, each cover may be activated for longer periods of time implying that the lifetime increases. This can be achieved by selecting grids with lower $TEL$.
\end{rem}

\begin{figure}[htbp]
\fbox{
\begin{minipage}{0.95\columnwidth}
\begin{algorithmic}

\STATE \textbf{Grid-Based Algorithm}
\vspace{2mm}
\STATE Generate collection of grids $\G$ based on shifting mechanism.

\REPEAT

\STATE Find Cover $C$ for grid next $G\in \G$.

\STATE Schedule $C$, i.e., reduce battery life of sensors in $C$ according to policy.

\IF{$G$ is tha last grid in $\G$} 

\STATE Restart from the first grid in $\G$

\ENDIF

\UNTIL no covers found for all grids in $\G$.

\end{algorithmic}
\end{minipage}}
\caption{Outline of Grid-Based algorithms.}

\label{fig:gridoutline}
\end{figure}

\subsubsection{Finding $\G$}
\label{sec:findgrids}

We describe a method of obtaining $\G$ by considering all possible placements of a tiling in $R$ such that the number of sensors which may cover any two tilings $G_{i}$ and $G_{j}$ is minimized. Let us examine the square and hexagonal grids separately.

\begin{figure}
  \centering
  \subfloat[\sl{Possible Square Grids are shown in Green, Blue and Red}]{
  \includegraphics[scale=0.4]{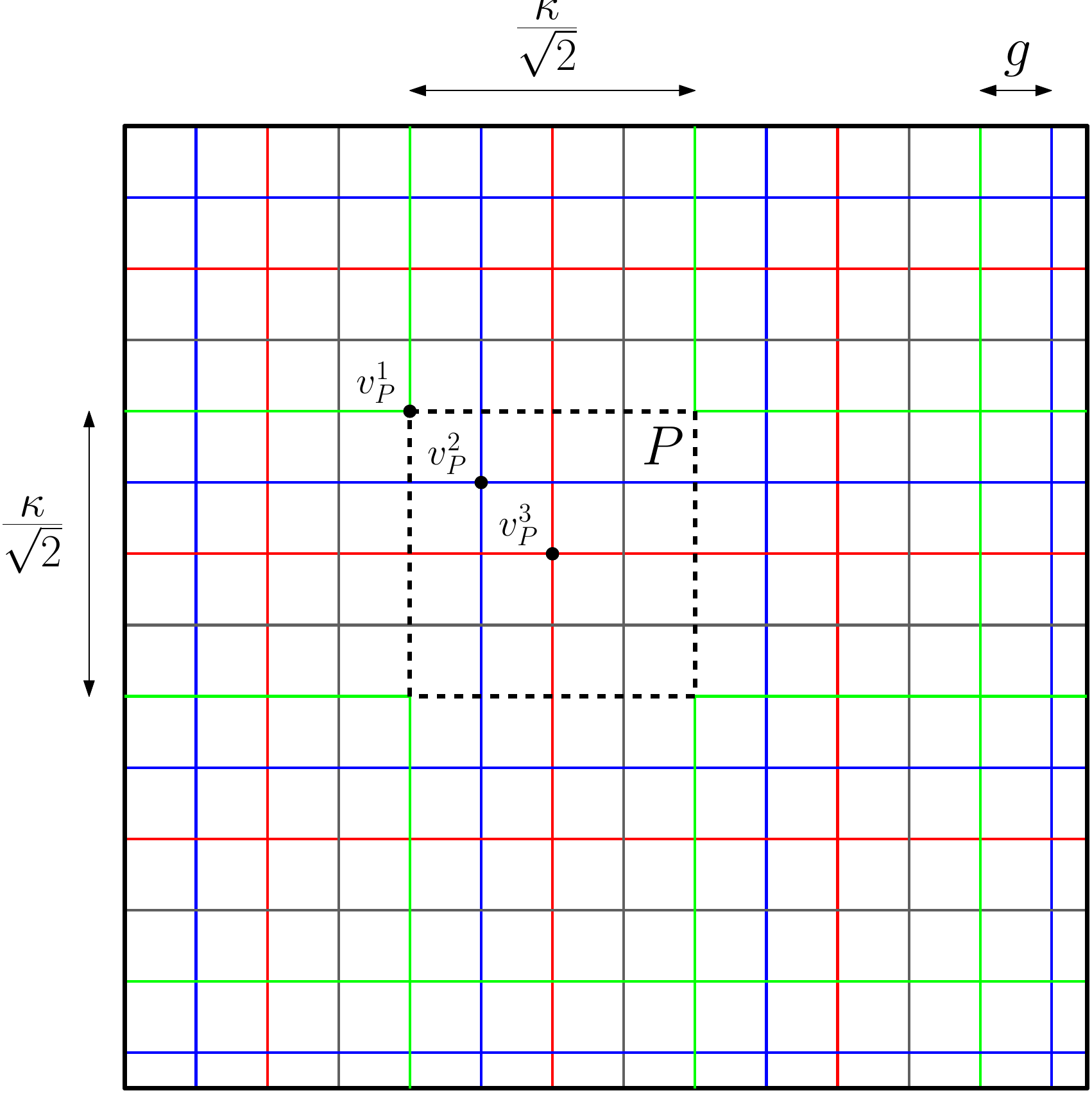}
  \label{fig:squareshift}
  }
  \hfil
  \subfloat[\sl{Possible Hexagonal Grids are shown in Green, Blue and Red}]{
  \includegraphics[scale=0.4]{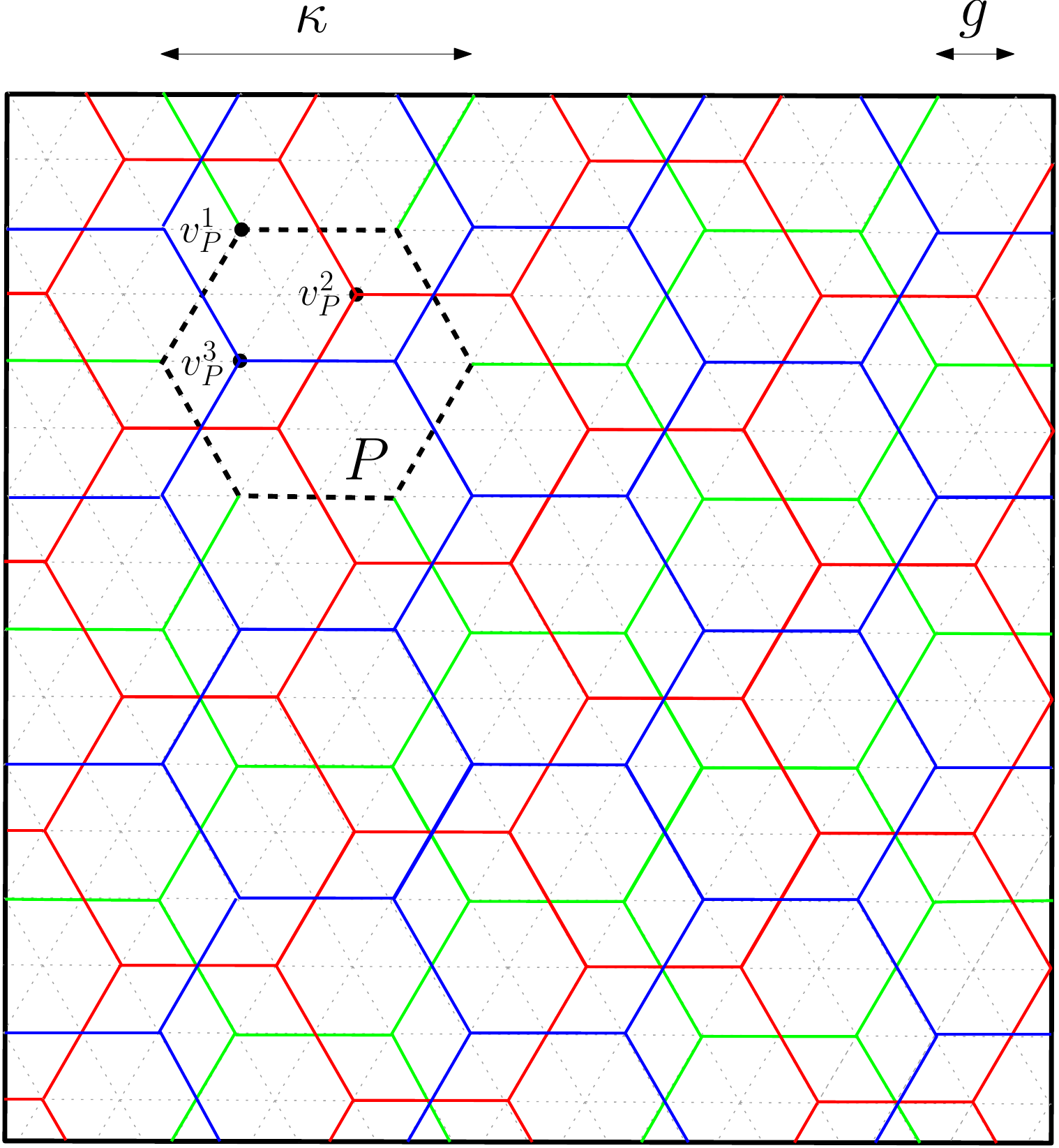}
  \label{fig:hexshift}
  }
  \caption{Shifting Mechanism for Grid-Based algorithms. In both the square and hexagonal cases, an underlying grid is used (square and triangular respectively). The points of shifting $v_{P}^{1},v_{P}^{2},v_{P}^{3}$ are shown inside the polygon $P$ shown with a dashed boundary. The value $g$ is the granularity. }
  \label{fig:shift}
\end{figure}

\paragraph*{Square Grid}In the case of the square grid, we start with an underlying grid of cells $g\times g$. Let this grid by $G_{U}$. Now, consider some square grid $G$ such thats its vertices are points in $G_{U}$, and some square $P$ in the tiling implied by $G$. Let $v_{P}$ be a vertex of $P$. Now, we may obtain other square grids by shifting $v_{P}$ into points of $G_{U}$ inside the original $P$ along the diagonal of $P$. The grids induced by $v_{P}$ will be the required $\G$ (see Figure \ref{fig:squareshift}).

\paragraph*{Hexagonal Grid}In the case of the hexagonal grid, we start with an underlying triangular grid $G_{U}$ where each triangle has side $g$. Let each grid $G$ be represented by some hexagon $P$ and let one of the vertices of $P$ be $v_{P}$. Now, we may obtain other hexagons by shifting $v_{P}$ to points of the triangular grid inside the original hexagon $P$ (see Figure \ref{fig:hexshift}). 

In both cases, the value $g$, which may be termed as \textbf{\emph{Granularity}} is key to ensuring that the covers are as disjoint as possible. In the case of the square grid, if the value $g$ is set to $2$, then it can be seen that a sensor can be part of at most two grids $G_{i}$ and $G_{j}$. This is because, for any two vertical(resp. horizontal) lines $l_{i}$ and $l_{j}$ part of grids $G_{i}$ and $G_{j}$ respectively, no sensor can cover both $l_{i}$ and $l_{j}$. Hence, a sensor can cover two grids $G_{i}$ and $G_{j}$ only at an intersection point of lines vertical line $l_{i}$ and horizontal line $l_{j}$ from grids $G_{i}$ and $G_{j}$ respectively. Similar values can be computed for the hexagonal grid also.

\subsubsection{Lifetime Guarantees}
\label{sec:gridlifetimeguar}

Setting the values of $g$ to $2$ and $x$ respectively, we obtain the following lemmas.

\begin{lemma}
\label{lem:Lsq}
  If a set of sensors $\S$ provides full coverage for $R$, then the
  lifetime of the system when using the square tiling to find
  individual covers is at least $\frac{\kappa}{2\sqrt(2)}$ when the schedule is uniform.
\end{lemma}
\begin{IEEEproof}
As seen in Section \ref{sec:findgrids}, setting the value of $g$ to $1$ implies that no sensor can be part of two vertical lines in two different grids. Hence, each sensor can be part of at most $2$ covers, one for which it intersects a vertical line and the other when it intersects the horizontal line. The number of shifts when $g=1$ is given by how many points are on the diagonal of a square of the tiling which are at least $\sqrt(2)$ apart. Hence, the number of shifts is $\frac{\kappa}{sqrt{2}}$. Since each sensor can be part of at most $2$ covers, we may activate each cover for $\frac{1}{2}$ units of time. Hence, the lifetime of the system is at least $\frac{\kappa}{sqrt{2}}\cdot \frac{1}{2} = \frac{\kappa}{2\sqrt{2}}$.
\end{IEEEproof}

The lifetime of the hexagonal grid can be similarly shown to be $\frac{\kappa \sqrt(3)}{4}$.
\commentswami{
\begin{lemma}
\label{lem:Lhex}
  If a set of sensors $\S$ provides full coverage for $R$, then the
  lifetime of the system when using the hexagonal tiling to find
  individual covers is at least $x$.  
\end{lemma}

\begin{IEEEproof}
 \commentswami{TODO}
\end{IEEEproof}}
This shows that the guaranteed lifetime of the hexagonal grid obtains better lifetime than the square grid.

\subsubsection{Approximation Scheme}
\label{sec:epsapprox}

\begin{figure}
  \centering
  \subfloat[\sl{Approximate Cover for $l$ when $l$ is an internal segment.}]{
  \includegraphics[scale=0.4]{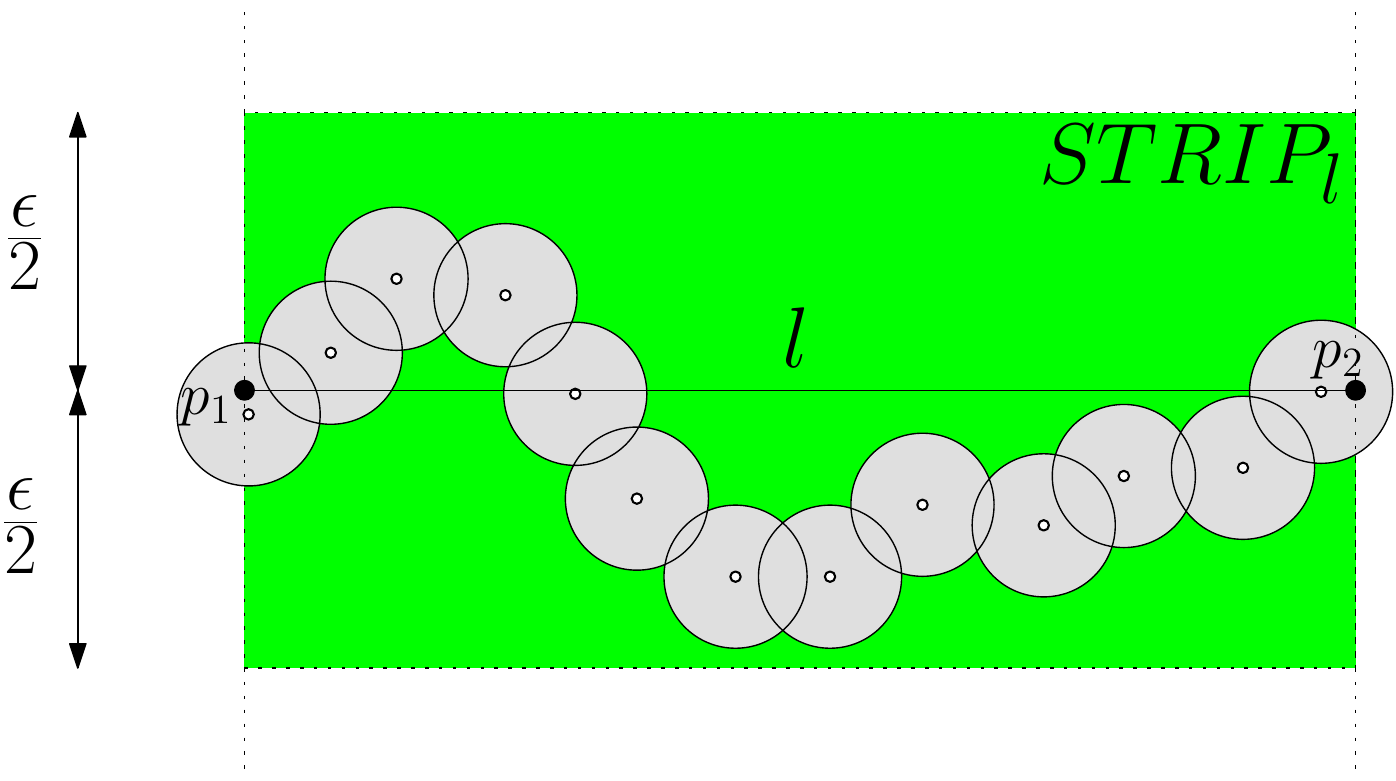}
  \label{fig:epsapproxint}
  }
  \hfil
  \subfloat[\sl{Approximate Cover for $l$ when $l$ intersects the boundary of $R$. The leftmost sensor is not restricted to cover $p_{1}$ but only to intersect the boundary of $R$.}]{
  \includegraphics[scale=0.4]{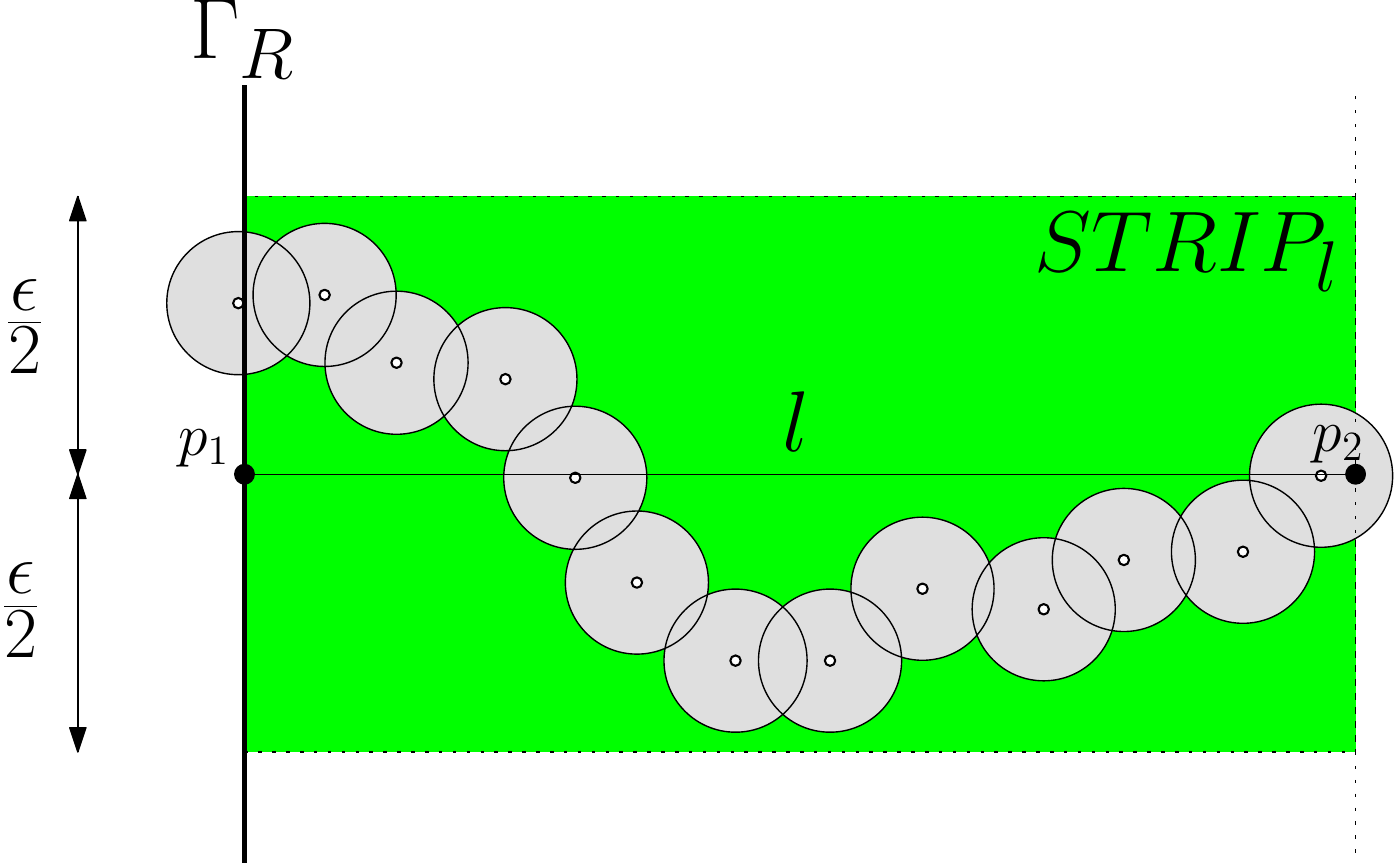}
  \label{fig:epsapproxbdry}
  }
  \caption{$\epsilon$-approximate cover shown for a line segment $l$ from $p_{1}$ to $p_{2}$. All sensors in the cover(shown as grey disks) are only restricted to lie within a strip of width $\epsilon$ which contains $l$. The requirements are (i) The sensors should be connected and (ii) If $l$ is an internal segment, the points $p_{1}$ and $p_{2}$ should be covered and if $l$ intersects the boundary of $R$, some sensor should intersect the boundary of $R$ within the strip. }
  \label{fig:epsapprox}
\end{figure}

Let the cover for a particular grid $G$ be $C$. Consider a line segment $l\in G$ and consider the sensors $C_{l}\subset C$ which cover $l$. Let $STRIP_{l}$ be a strip of width $\epsilon$ containing $l$ and equidistant from $l$ on either side and let the boundary of $R$ be $\Gamma_{R}$. In the approximation scheme, $C_{l}$ need not cover the whole of $l$. Instead,it is only required to cover $STRIP_{l}$ (see Figure \ref{fig:epsapprox}). Specifically, the requirements for the cover $C_{l}$ differ based on whether $l$ intersects the $\Gamma_{R}$,

\begin{enumerate}
\item $C_{l}$ should be connected
\item If $l$ is an internal segment, then $C_{l}$ must cover both endpoints of $l$ (see Figure \ref{fig:epsapproxint}). Otherwise, if one or both ends of $l$ intersect $\Gamma_{R}$, then $C_{l}$ must cover some portion of $\Gamma_{R}$ within $STRIP_{l}$ (see Figure \ref{fig:epsapproxbdry}).
\end{enumerate}

This scheme yields covers where there may be curves whose endpoints are at most $(1+\epsilon)\kappa$ away from each other. In other words, the cover is a $(1+\epsilon)\kappa$-weak cover.

In Section \ref{sec:results}, we see that this yields great dividends in lifetime.

\subsection{Random Seeds}
\label{sec:randoverview}

The random seeds technique involves selecting at random a subset $S\in \S$ as
seeds and turning off sensors surrounding these seeds to form holes as
large as possible while maintaning the $\kappa$-weak property. The
algorithms using which the sensors are turned off depends on whether
GPS is available and is described in Section \ref{sec:findcover}.

\commentswami{I am not sure about the motivation for the Random Seeds}

\section{Finding a Cover $C$ and an activation time $\delta$}
\label{sec:findcover}

In this section, we describe several heuristics to find a cover $C$ for a given grid $G$ and an activation time $\delta$ for $C$.

\begin{defn}
  A \textbf{\emph{Minimal $\kappa$-weak cover}} is defined as a $\kappa$-weak cover $C$ for $R$ such that $C\setminus s$ for any $s\in C$ is not a $\kappa$-weak cover.
\end{defn}

In the following algorithms, we are interested in finding minimal covers always. This is because, for any cover which is not minimal, the redundant sensors would worsen the second criteria described in Remark \ref{rem:lifetimecriteria}.

\subsection{Finding a cover for the Grid-Based Algorithm}
\label{sec:findcovergrid}

In each of the following algorithms, we are given as input a grid $G=\{l_{1},l_{2},\dots,l_{k}\}$. Let the graph $A$ be the unit disk graph of sensors $S$, i.e., we have a vertex corresponding to each sensor and an edge for every two sensors whose ranges intersect. $A$ is unweighted and undirected.

\subsubsection{Min-Max Heuristic}
\label{sec:greedy}

The Min-Max Heuristic is described by Balasubramaniam et.al. \cite{ravi-coverage} for full coverage. The idea is that at each stage, the sensors are arranged in decreasing order of their battery life. Then, the algorithm checks whether the removal of sensor $s\in S$ (selected in order) violates the cover. If yes, it is made active and otherwise, made inactive. For more details, refer to \cite{ravi-coverage}.

In our case, instead of checking whether removal of a sensor violates full coverage, we check if removal of a sensor violates the coverage of the grid $G$. Let $s$ be the sensor under consideration and let $l$ be a line which intersects $s$ whose endpoints are $p_{1}$ and $p_{2}$. Also, let $A_{l}=(S_{l},E_{l})$ be the subgraph of $A$ induced by sensors $S_{l}$ which intersect $l$. Now, removal of sensor $s$ may violate coverage only for such lines. Hence, when we remove $s$, we check if all lines $l$ are covered. This is done by checking if the remaining sensors in $S_{l}$ are connected and also whether $p_{1}$ and $p_{2}$ are covered. In the case of the approximation scheme, the second requirement is modified as described in Section \ref{sec:epsapprox}. 

\subsubsection{Breadth-First-Search (BFS)}
\label{sec:bfs}

\begin{figure}[htbp]
\fbox{
\begin{minipage}{0.95\columnwidth}
\textbf{Breadth-First-Search Algorithm}
\begin{algorithmic}[1]

\vspace{2mm}
\STATE Given a grid $G=\{l_{1},l_{2},\dots,l_{k}\}$ and graph $A$ from $S$ and a paramater $\epsilon$ such that $0\leq \epsilon \leq 1$.

\STATE Let $\Gamma_{R}$ be the boundary of $R$.

\STATE Let $\mathcal{B}=\{b_{1},b_{2},\dots,b_{m}\}$ be the set of possible battery lives of sensors $S$.

\STATE Perform binary search on $\mathcal{B}$ to find the maximum lifetime $b_{j}$ such that a cover is found for $G$ with battery life $b_{j}$. At each step of the binary search when $b$ is the current battery, do the following steps.

\FOR{each line $l_{i}\in G$ with endpoints $p_{1}$ and $p_{2}$}

\STATE Let $STRIP_{l_{i}}$ be the strip of width $\epsilon\cdot \kappa$ around $l_{i}$ equidistant on either side.

\STATE Let $S_{l}\subset S$ be the set of sensors $s$ such that $D_{s}\cup STRIP_{l_{i}}\neq \phi$ and $B(s)\geq b$.

\STATE Let $S_{s}$ (resp. $S_{d}$) be the set of sensors coverin $p_{1}$ (resp. $p_{2}$). For the approximation scheme, if $p_{1}$ (similarly $p_{2}$) lies on $\Gamma_{R}$, let $S_{s}$ (similarly $S_{d}$) be the set of sensors which intersect the portion of $\Gamma_{R}$ inside $STRIP_{l_{i}}$.

\STATE Augment $A_{l_{i}}$ with two vertices $s$ and $d$ and add edges from $s$ to $S_{s}$ and $d$ to $S_{d}$.

\STATE Find the shortest path $\pi$ from $s$ to $d$.

\STATE $C_{l}=\{s:s\in pi\}$.

\ENDFOR

\STATE $C=\displaystyle{\bigcup_{l\in G} C_{l}}$.

\end{algorithmic}
\end{minipage}}
\caption{Steps of the Breadth-First-Search Algorithm}

\label{fig:bfssteps}
\end{figure}

\begin{figure}
  \centering
  \includegraphics[scale=0.7]{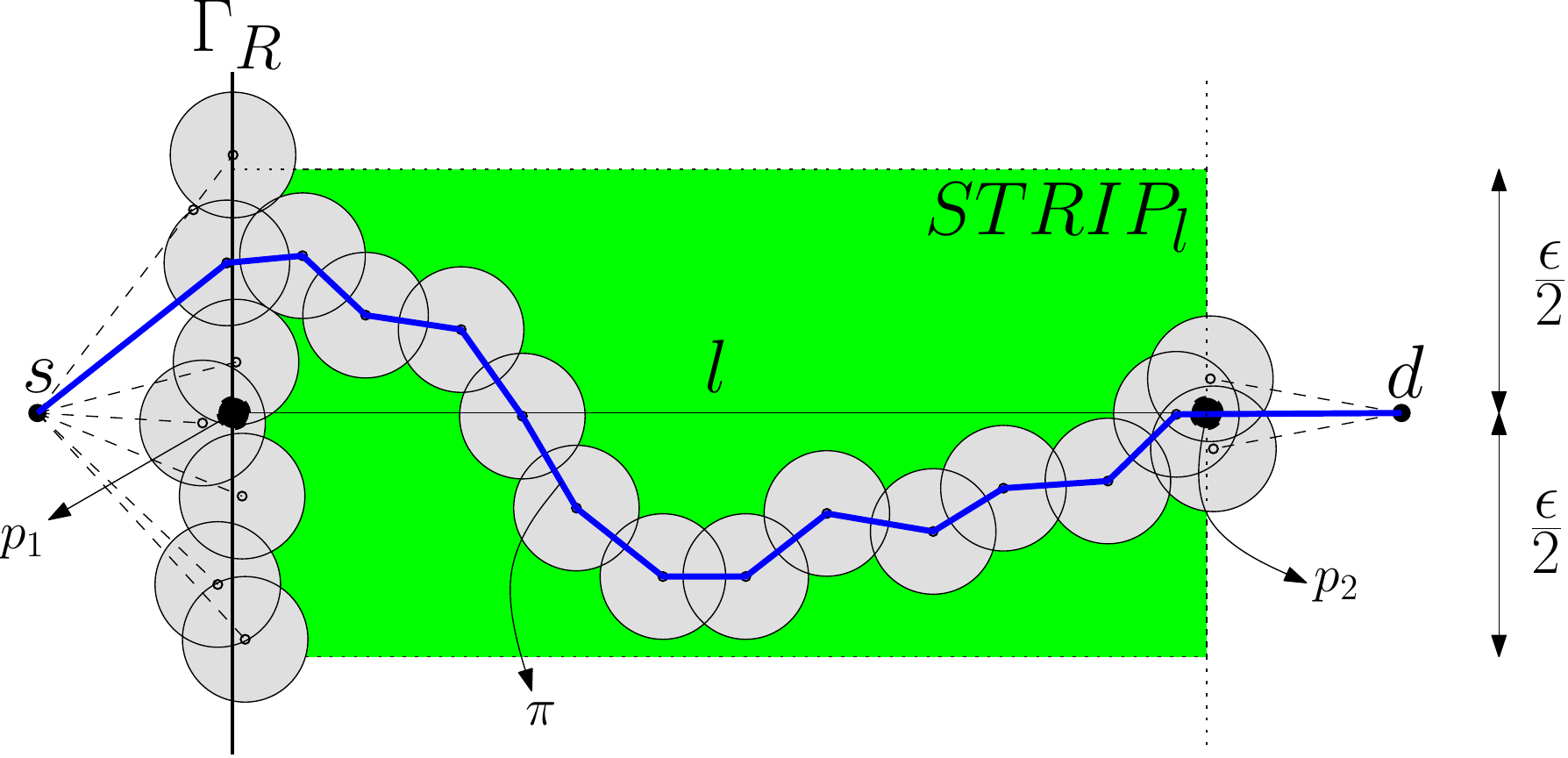}
  \caption{Example of augmented graph and the result of the Breadth-First-Search Algorithm. Here, $p_{1}$ lies on $\Gamma_{R}$ and $p_{2}$ is an internal point. Two additional vertices $s$ and $d$ are added. $s$ is connected to all sensors intersecting $\Gamma_{R}$ within $STRIP_{l}$. $d$ is connected to all sensors within $STRIP_{l}$ covering $p_{2}$. The result of the BFS algorithm is the path $\pi$ shown as BLUE fat lines. All sensors in $\pi$ form the cover $C_{l}$.}
  \label{fig:bfs}
\end{figure}

We describe the BFS algorithm for the approximation scheme with some parameter $\epsilon$ such that $0\leq \epsilon\leq 1$. In the exact case, simply set $\epsilon$ to $0$. $\Gamma_{R}$ and $STRIP$ are defined as in Section \ref{sec:epsapprox}. Now, for some line $l\in G$ with endpoints $p_{1}$ and $p_{2}$, consider some set of sensors $S_{l}$ which intersect $STRIP_{l}$. Let $A_{l}$ be the subgraph of $A$ induced by $S_{l}$. Now, we define two sets of sensors $S_{s}$ and $S_{d}$ as follows. If $l$ is an internal segment, then $S_{s}$ (resp. $S_{d}$) be the set of sensors covering $p_{1}$ (resp. $p_{2}$). If $p_{1}$ (similarly $p_{2}$) lie on $\Gamma_{R}$, then, let $S_{s}$ (similarly $S_{d}$) be the set of sensors which intersect the portion of $\Gamma_{R}$ inside $STRIP_{l}$. Figure \ref{fig:bfs} shows how $S_{s}$ and $S_{d}$ are generated.

Let $\mathcal{B}=\{b_{1},b_{2},\dots,b_{m}\}$ be the set of possible battery lives of $S$. We perform binary search on $\mathcal{B}$ to find the cover with maximum battery life. At an intermediate stage of this search, when $b$ is the current battery, the set $S_{l}$ is found by finding sensors intersecting $l$ which have battery life at least $b$. This is done for each $l\in G$. Then, for each $A_{l}$, the augmentation with two vertices $s$ and $d$ is performed as described above. Now, all we need to do is find if there is a path from $s$ to $d$ for every $l\in G$ (see Figure \ref{fig:bfs}). If such paths exist, the sensor constituting the paths will form the minimal cover $C$. See Figure \ref{fig:bfssteps} for the steps of the algorithm.

The time complexity of the BFS algorithm is $O(m\cdot \frac{L}{\kappa}\cdot n)$, where $m$ is the number of unique battery values. This is because there are at most $m$ times we may perform the search for the shortest path and at most $\frac{L}{\kappa}$ lines and each search takes $O(n)$ time. This is better than the Min-Max heuristic because, for the Min-Max heuristic, the check when each sensor is removed may take $O(n)$ time leading the overall complexity to be $O(n^{2})$. In practice, the values $\frac{L}{\kappa}$ and $m$ are constant and hence the time complexity of the BFS algorithm is only $O(n)$.

\subsubsection{Maximum Flow}
\label{sec:maxflow}

Here, we describe a maximum flow algorithm for the square grid case. Given a grid, the maximum flow algorithm finds a number of covers $C$ and activation times $\Delta$. The algorithm has two parts: (i) First, we find the maximum flow and (ii) Second, we find individual paths carrying flow where each one is thought of as a cover and the time for the cover is the flow through this path.

\paragraph*{Finding Maximum Flow via Linear Programming}

\begin{figure}
  \centering
  \subfloat[\sl{Augmentation of $A_{G}$ with additional vertices $s,d$ and $\mu$ and edges from and to $\mu$, from $s$, to $d$ and between boundary sensors intersecting adjacent lines. Also shown are the four types of regions in $R$.}]{
\centering
  \includegraphics[scale=0.37]{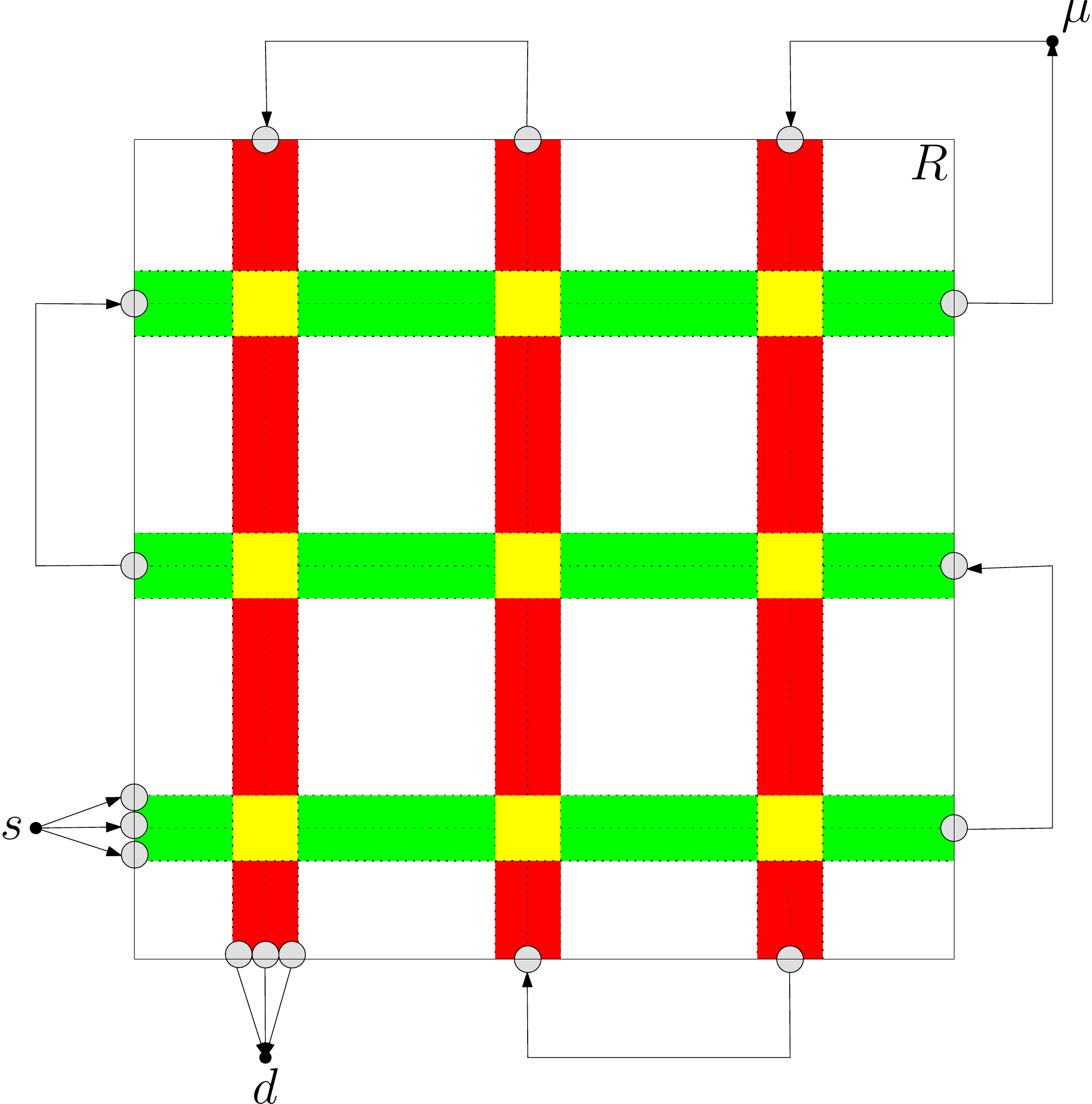}
  \label{fig:LPoverview}
  }
  \hfil
  \subfloat[\sl{Replacement of a single sensor $s_{i}$ into $s_{i,in}$ and $s_{i,out}$. All sensors $\{s_{1},s_{2},\dots,s_{k}\}$ adjacent to $s_{i}$ have been replaced and the corresponding edges are split into two edges from $\{s_{1,out},s_{2,out},\dots,s_{k,out}\}$ to $s_{i,in}$ and vice versa. The capacities of these edges is unconstrained. The capacity of the internal edge is the battery $B(s_{i})$.}]{
\centering
  \includegraphics[scale=0.5]{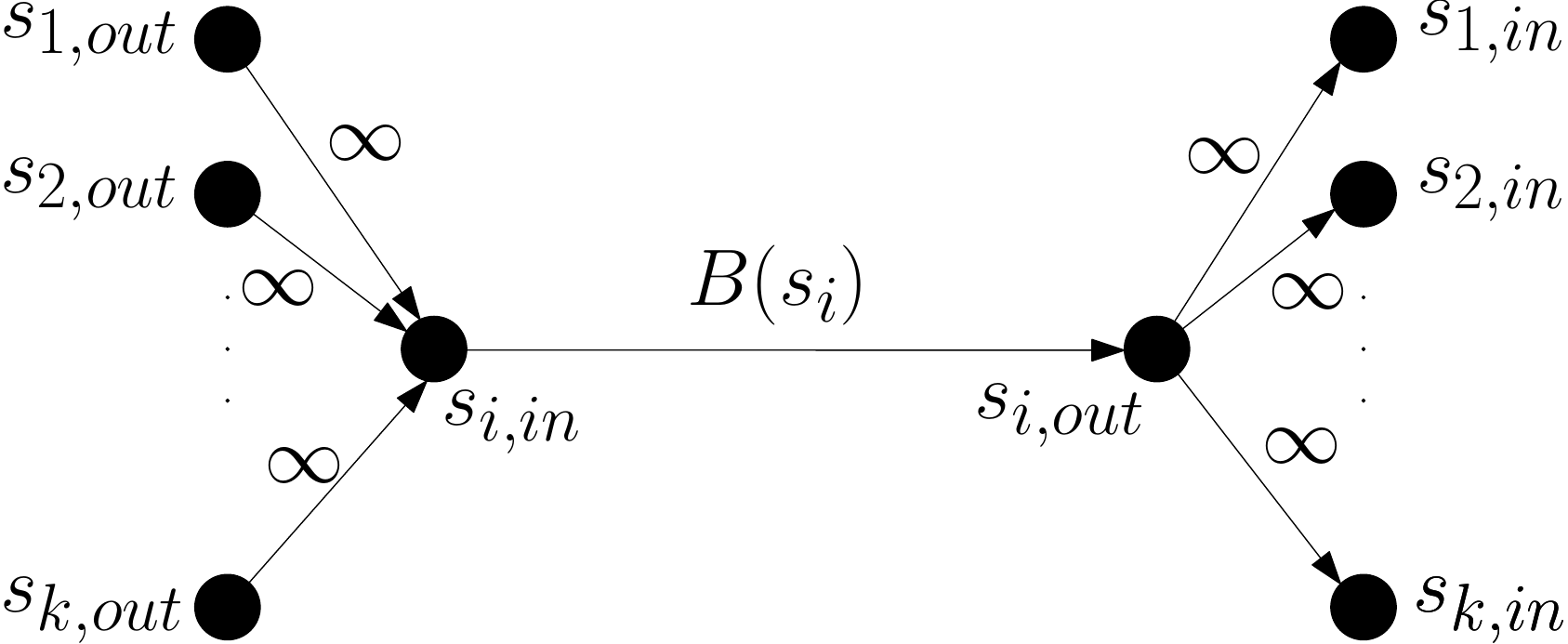}
  \label{fig:Lpsinglevertex}
  }
  \caption{Formation of directed graph $A_{LP}$ from $A_{G}$.}
  \label{fig:LP}
\end{figure}

Let the sensors which intersect the grid $G$ be $S_{G}$ and let $A_{G}$ be the subgraph of $A$ induced by $S_{G}$.  We form a directed graph $A_{LP}$ from $A_{G}$ as follows. Each sensor $s_{i}$ is replaced by two vertices $s_{i,in}$ and $s_{i,out}$ with an edge $(s_{i,in},s_{i,out})$. For each edge $(s_{i},s_{j})$ of $A_{G}$, we have two edges $(s_{i,out},s_{j,in}),(s_{j,out},s_{i.in})$ in $A_{LP}$. 

The region $R$ is divided into $4$ types (See Figure): (i) Vertical Only (marked red in Figure), (ii) Horizontal Only (marked green), (iii) Mixed (marked yellow) and (iv) Non-Grid (marked Gray). We define two types of flow in $A_{LP}$, hence, it is a multi-commodity flow problem. On each edge $(s_{i,out},s_{j,in})$ in $A_{LP}$, we can have horizontal (resp. vertical) flow only if both $s_{i}$ and $s_{j}$ intersect the horizontal(resp. vertical) region. 

Let the horizontal and vertical lines be ordered as $G_{H}=\{l_{1}^{h},l_{2}^{h},\dots,l_{k}^{h}\}$ and $G_{V}=\{l_{1}^{v},l_{2}^{v},\dots,l_{k}^{v}\}$. Let $S_{i,R}^{H}$ (resp. $S_{i,L}^{H}$) denote the set of sensors which intersect $STRIP_{l_{i}^{h}}$ and the right (resp. left) boundary of $\Gamma_{R}$ and let $S_{i,B}^{V}$ (resp. $S_{i,T}^{V}$) denote the set of sensors which intersect $STRIP_{l_{i}^{v}}$ and the bottom (resp. top) boundary of $\Gamma_{R}$. For each pair of sensors $s_{a},s_{b}$ such that $s_{a}\in S_{i,R}^{H}$ and $s_{b}\in S_{i+1,R}^{H}$ for some $i$, we add the edge $(s_{a,out},s_{b,in})$. Similarly, this is done for sensors in $S_{i+1,L}^{H}$ to $S_{i+2,L}^{H}$ and so on. We also add similar edges for the vertical lines. Finally, we add three vertices $s,d$ and $\mu$ to $A_{LP}$ and edges are added as follows:

\begin{enumerate}
\item For every sensor $s_{a}\in S_{1,L}^{H}$, we add the edge $(s,s_{a,in})$.
\item For every sensor $s_{b}\in S_{k,T}^{V}$, we add the edge $(s_{b,out},d)$.
\item For every sensor $s_{c}\in S_{k,R}^{H}$, we add the edge $(s_{c,out},\mu)$.
\item For every sensor $s_{c}\in S_{1,B}^{V}$, we add the edge $(\mu,s_{d,in})$.
\end{enumerate}

See Figure \ref{fig:LP} for an illustration of these additional edges. There are no constraints on the capacity of edges between two sensors. On each horizontal edge and on edges from $s$, only horizontal flow is permitted and on each vertical edge and on edges to $d$, only vertical flow is permitted. The internal edge $(s_{i,in},s_{i,out})$ has capacity equal to the battery life of $s_{i}$, $B(s_{i})$ and the flow permitted depends on which region $s_{i}$ intersects. The purpose of the vertex $\mu$ is to convert horizontal flow into vertical flow. Hence, edges into $\mu$ have horizontal flow and edges outward from $\mu$ have vertical flow. For any vertex other than $s$ and $d$, the incoming flow is equal to the outgoing flow and the flow out of $s$ is equal to the flow into $d$.

Let the flow on a horizontal (resp. vertical) edge $(u,v)$ be $x_{u,v}$ (resp. $y_{u,v}$). Then, linear program for finding maximum flow is given in Figure \ref{fig:LPform}.

\begin{figure}[htbp]

\begin{minipage}{0.95\columnwidth}
\begin{IEEEeqnarray}{lrCl}
\textbf{Maximize } & \sum_{s_{i}} x_{s,s_{i}} && \nonumber \\
\textbf{subject to: } &  \forall s_{i}\text{ } \sum_{s_{j}} x_{s_{j},s_{i}} & = & \sum_{s_{j}} x_{s_{i},s_{j}} \\
& \forall s_{i}\text{ } \sum_{s_{j}} y_{s_{j},s_{i}} & = & \sum_{s_{j}} y_{s_{i},s_{j}} \\
& \forall s_{i}\text{ } \sum_{s_{j}} x_{s_{i},s_{j}} + \sum_{s_{j}} y_{s_{i},s_{j}} & \leq & B(s_{i}) \\
& \sum_{s_{i}} x_{s_{i},\mu} & = & \sum_{s_{i}} y_{\mu,s_{i}} \\
& \sum_{s_{i}} x_{s,s_{i}} & = & \sum_{s_{i}} y_{s_{i},d}
\end{IEEEeqnarray}

\end{minipage}
\caption{Linear Program for finding Maximum Flow}
\label{fig:LPform}

\end{figure}

\paragraph*{Finding Invidual Covers}

The solution to the linear program gives a flow on each edge in $A_{LP}$ and the maximum flow  corresponds to a set of covers $C_{G}$ for $G$ and an activation time $\delta_{C}$ for each cover $C\in C_{G}$. Each cover $C$ corresponds to a path from $s$ to $d$ with flow $\delta_{C}$. Hence, $C$ may be found by finding the augmenting path from $s$ to $d$ in $A_{LP}$. This is done by greedily exploring from $s$ along edges containing maximum flow. Once a path $pi$ is found, the corresponding cover $C$ is the set of sensors $pi$. $\pi$ is then deaugmented, i.e., the flow along every edge in $\pi$ is reduced by the flow along $pi$. Note that these paths may contain loops, in which case, every loop is shortened to get the flow assignment to each edge. Once the deaugmentation is complete, the cover may be activated based on the Non-Uniform activation policy described in Section \ref{sec:actpolicy}. The Uniform activation policy does not apply to this method since the optimal solution to the LP gives the activation time for each cover in order to get maximum flow. The Non-Preemptive activation policy also does not apply since the optimal solution to the LP may cause a sensor to be activated for two different covers in $C_{G}$. This process is repeated until no more paths are found. 

\subsection{Finding a cover for the Random Seeds Algorithm}
\label{sec:randseeds}

The random seeds algorithm is designed to operate when GPS is not available, i.e., the sensors do not know their exact location. During the course of the algorithm, nodes are labeled as ``Boundary'', ``Deactivated''. Initially, all the nodes are unlabeled. The algorithm works by randomly picking a set of $k$ seeds and performing Breadth First Search from each of them. The search proceeds through unlabeled nodes and stops when a node which is either marked ``Boundary'' or which is at exactly $\kappa$ hops is reached, i.e., the search is not continued from such nodes. Also when a node is reached by searches from two or more seeds simultaneously, it is marked as ``Boundary'' and the search is not continued further. At the end, those sensors which are exactly $\kappa$ hops are marked as ``Boundary'' and others found during the search are marked ``Deactivated''. This is then repeated for all the seeds. 

In our preliminary simulations, the random seeds algorithm does not achieve comparable lifetime compared to the other methods. However, this is the only algorithm which does not require GPS and hence, is worth looking into to make improvements.

\commentswami{
\begin{figure}[htbp]
\fbox{
\begin{minipage}{0.95\columnwidth}
\begin{algorithmic}

\STATE \textbf{Random Seeds Algorithm}
\vspace{2mm}
\STATE Generate the Adjacency Graph $\G$ of $S$ where two sensors are neighbors if their ranges overlap.

\STATE All sensors $s\in S$ where battery $B(s)>0$ are unmarked.

\REPEAT

\STATE Randomly pick a subset of sensors $S_{seeds}$ for use as seeds among unmarked sensors.

\FOR{each seed $s\in S_{seeds}$}

\STATE Execute $BFS$ from $s$.

\FOR{every node $u$ visited during $BFS$}

\IF{$|u-s|\geq \kappa$ or $u$ is marked ``BOUNDARY''}

\STATE Mark $u$ as a ``BOUNDARY'' if not already marked.

\STATE Stop $BFS$ execution at $u$, i.e., do not explore to neighbors of $u$.

\ELSE

\STATE Mark $u$ as ``OFF''

\ENDIF

\ENDFOR

\ENDFOR

\UNTIL two iterations give the same set of sensors marked ``BOUNDARY''

\STATE All sensors marked ``BOUNDARY'' are part of cover $C$.

\STATE Activate cover $C$ based on activation policy.

\end{algorithmic}
\end{minipage}}
\caption{Steps of the Random Seeds algorithm}

\label{fig:algrandseeds}
\end{figure}
}

\subsection{Finding an activation time $\delta$ for a cover $C$}
\label{sec:actpolicy}

We describe three methods to assign activation time $\delta$ for a cover $C$. These three methods correspond to the three types of schedules described in Section \ref{sec:form}. 

\paragraph*{Uniform}

In the Uniform activation policy, each cover is assigned the same amount of time for which to be active. The method in which this is assigned follows the description of the Min-Max heuristic in \cite{ravi-coverage}.

\begin{defn}
  The \textbf{\emph{Load}} of a sensor, denoted by $load(s)$ is defined as the number of covers in which $s$ participates.
\end{defn}

We assign each sensor $s$ a maximum value for $load(s)$ at the start and at the discovery of each cover, we reduce the battery of the sensor by $\frac{1}{m}$. Hence, after $m$ covers, the battery will be depleted. In this manner, we may find the maximum number of covers such that the maximum load of every sensor is as assigned. Note that this policy will not be applicable in the case of the maximum flow algorithm since the maximum flow algorithm gives a solution in which the durations of the individual covers may not be equal.

\paragraph*{Non-Uniform}

In this policy, after finding each cover $C$, we find the sensor $s\in C$ which has minimum battery $b$. Then, we reduce the batteries of all the sensors in $C$ by a value $d\cdot \displaystyle \min_{s\in C} B(s)$, where $d$ is a user-defined parameter termed as the \textbf{\emph{decay coefficient}}.

\paragraph*{Non-Preemptive}

In the Non-Preemptive case, once a cover is found, it is activated for the entire battery life of the sensors.

\section{Experimental Results}
\label{sec:results}

\begin{figure}[t]

\centering
\subfloat[\sl{$\epsilon=0$}]{
\includegraphics[scale=0.35]{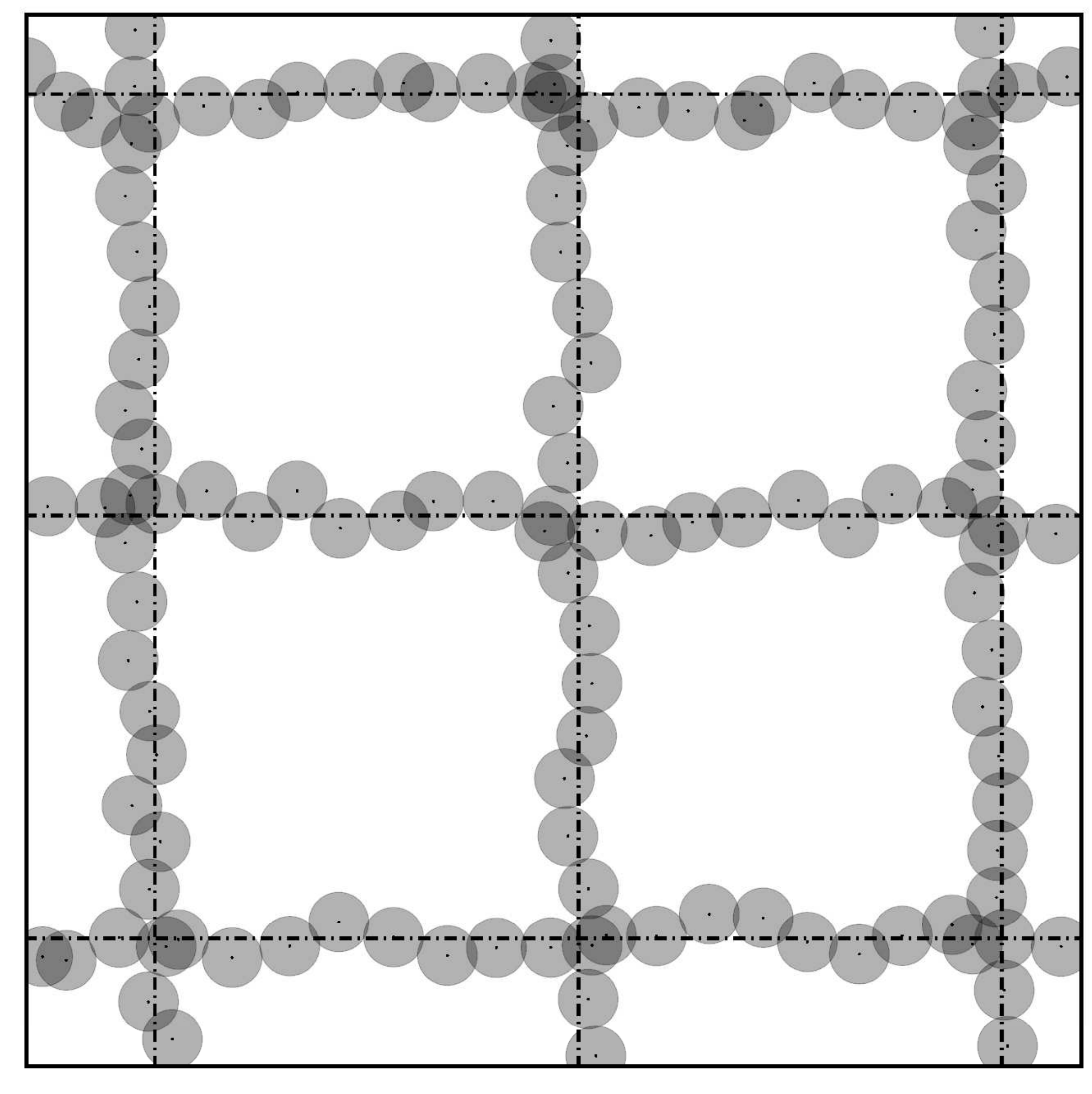}
\label{fig:cov0}
}
\hfil
\subfloat[\sl{$\epsilon=0.1$}]{
\includegraphics[scale=0.35]{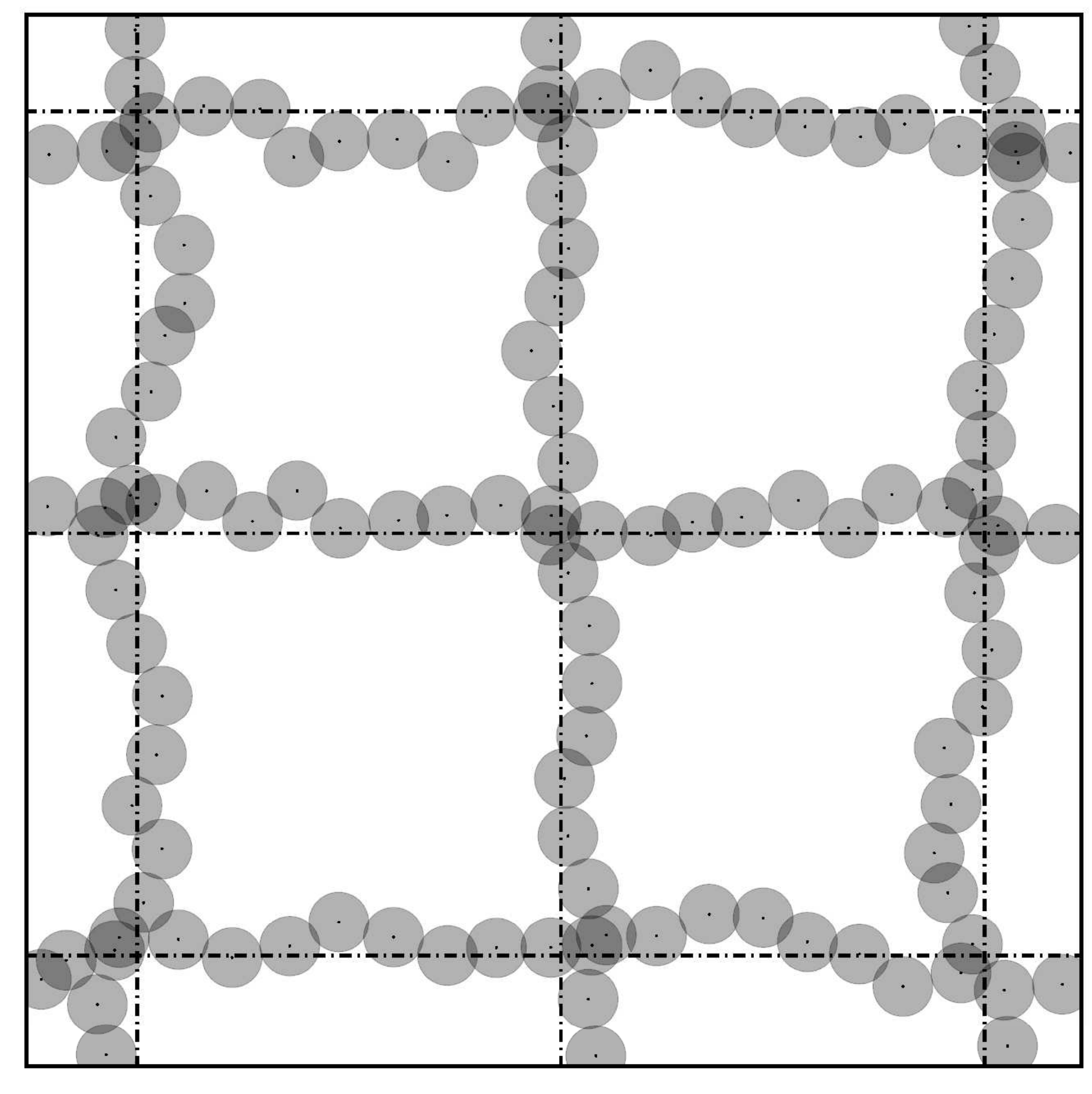}
\label{fig:cov1}
}
\hfil
\subfloat[\sl{$\epsilon=0.2$}]{
\includegraphics[scale=0.35]{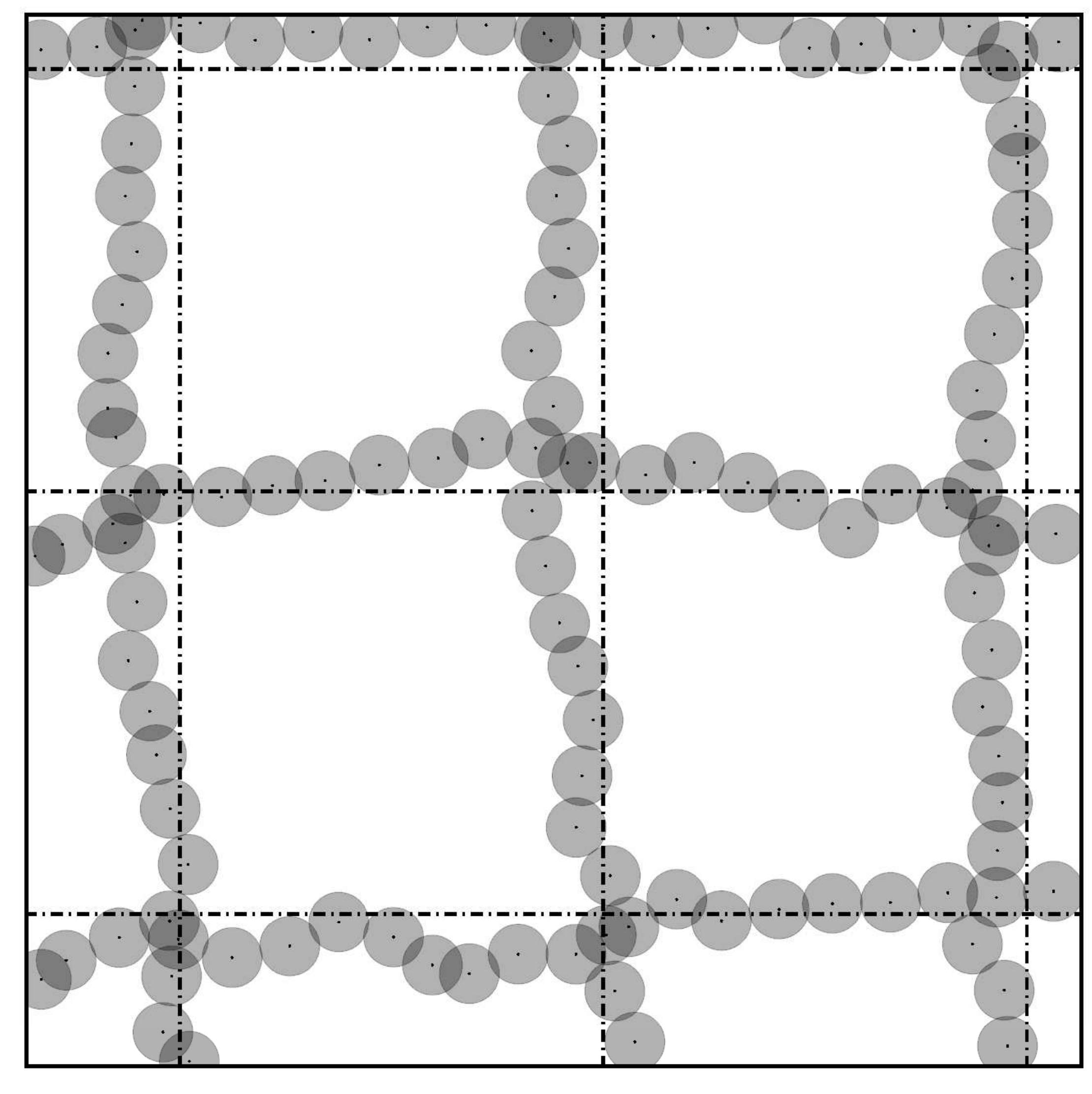}
\label{fig:cov2}
}
\hfil
\subfloat[\sl{$\epsilon=0.3$}]{
\includegraphics[scale=0.35]{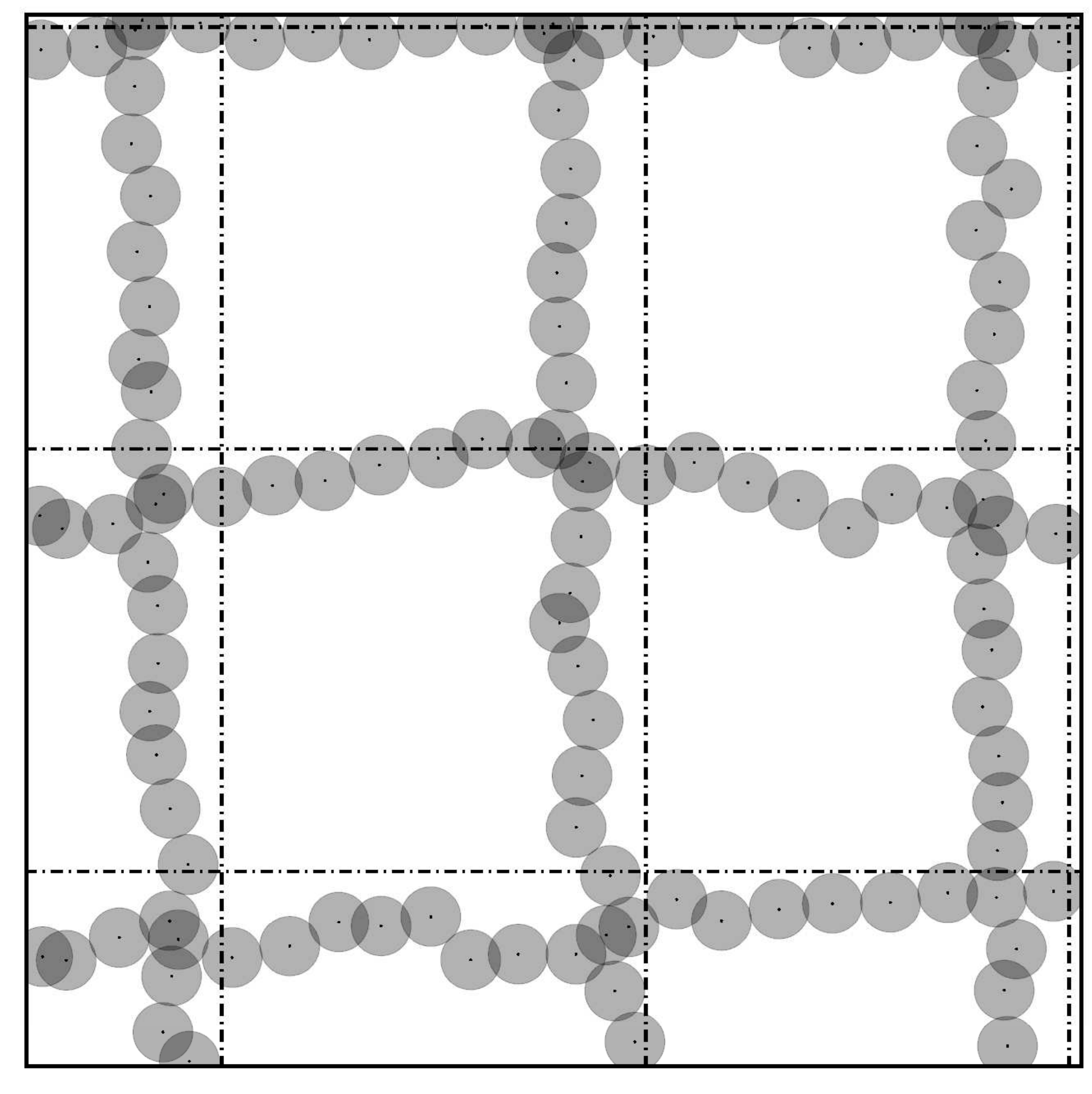}
\label{fig:cov3}
}
\caption{}
\label{fig:covers}
\end{figure}

In our simulations, sensors are placed in a $50\times 50$ region $R$. $R$ is then divided into cells of size $1\times 1$ and sensors are placed randomly inside the cells according to a spatial poisson process of some \emph{Intensity}. We generated $50$ such sets of sensors of intensities $1$ and $2$ and the mean of the lifetime values was calculated. The algorithms \textbf{Min-Max} and \textbf{BFS} algorithms were simulated with all three activation policies in order to compare them with the square grid. In the case of the hexagonal grid, we simulated the BFS algorithm with uniform activation policy to compare its performance with the square grid. The LP was simulated with non-uniform activation policy for the square grid. Each of these simulations was performed with $\kappa=\{10,20\}$ and $\epsilon=\{0,0.1,0.2,0.3\}$. The granularity of the shifts was set to $g+\epsilon$, where $g$ is set to the values from Section \ref{sec:findgrids}, since we want to capture as disjoint sensors as possible with each shift. Figure \ref{fig:covers} shows four example covers for the square grid for $\epsilon=\{0,0.1,0.2,0.3\}$. It can be seen that as $\epsilon$ increases, the sensors deviate further and further from the lines of the grid.

\begin{figure}[p]
\begin{minipage}[b]{0.5\linewidth}
  \centering
  \subfloat[\sl{Uniform}]{
  \includegraphics[scale=0.35]{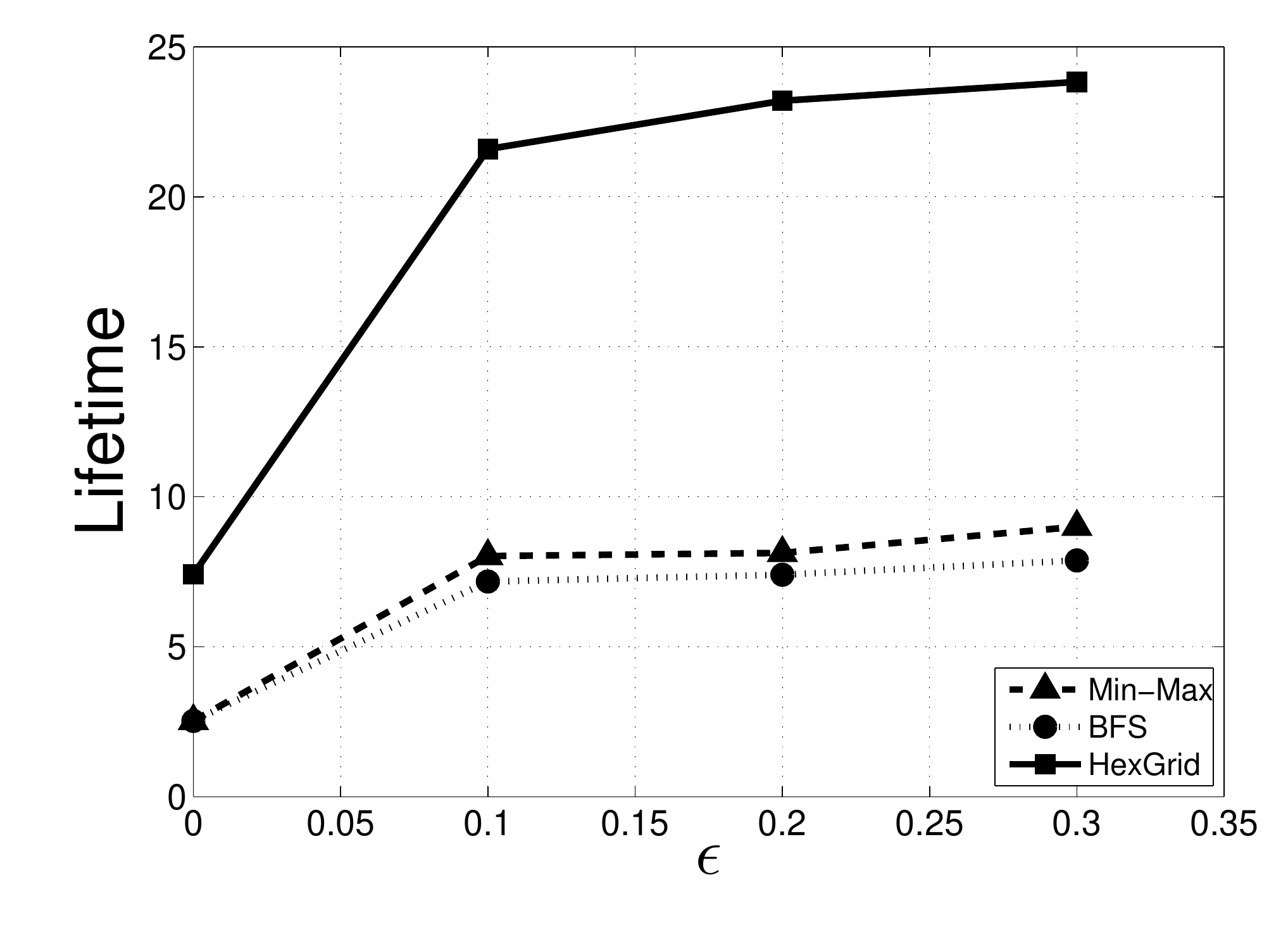}
  \label{k10i1un}
  }
  \hfil
  \subfloat[\sl{Non-Uniform}]{
  \includegraphics[scale=0.35]{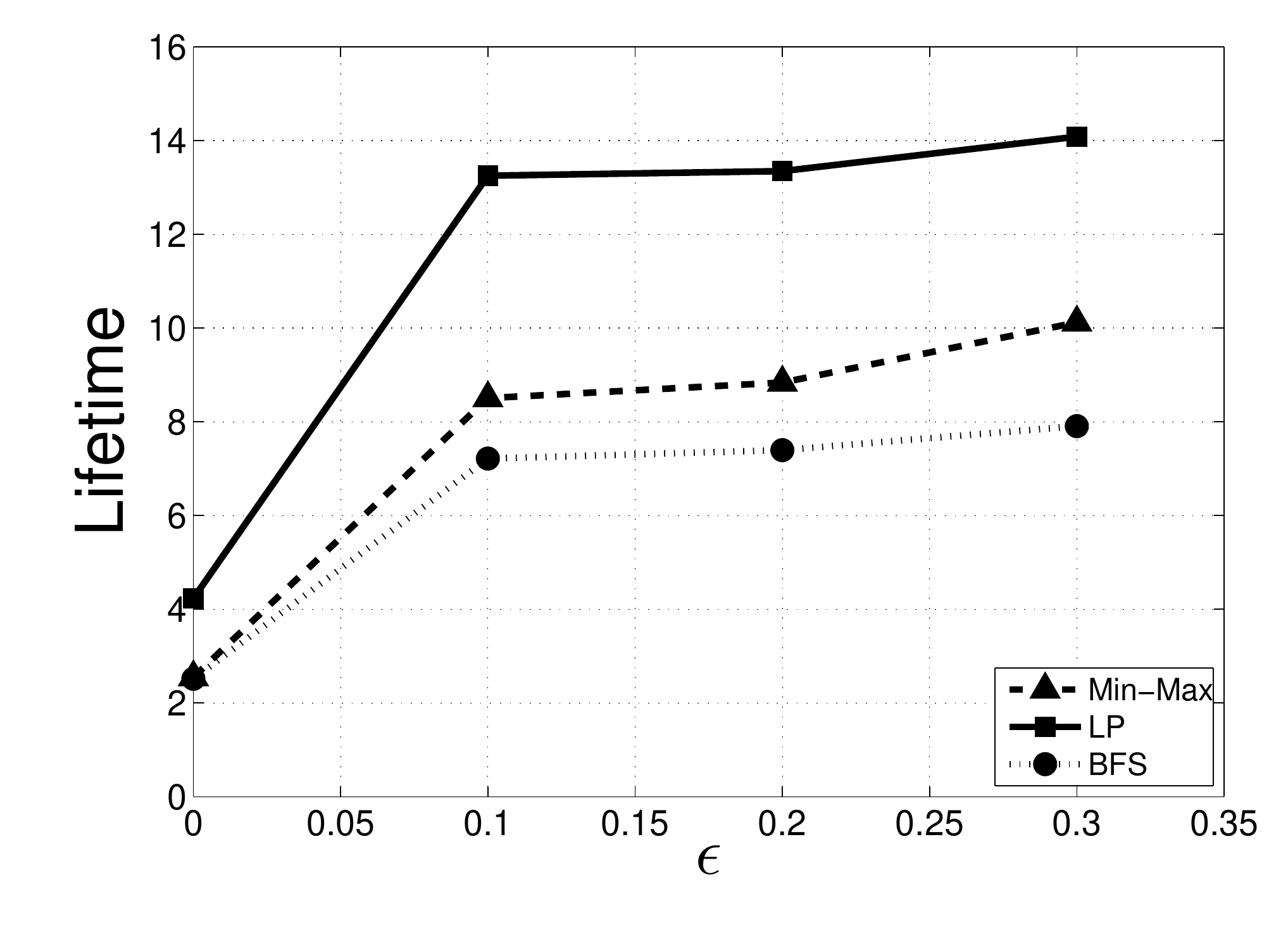}
  \label{k10i1de}
  }
  \hfil
  \subfloat[\sl{Non-Preemptive}]{
  \includegraphics[scale=0.35]{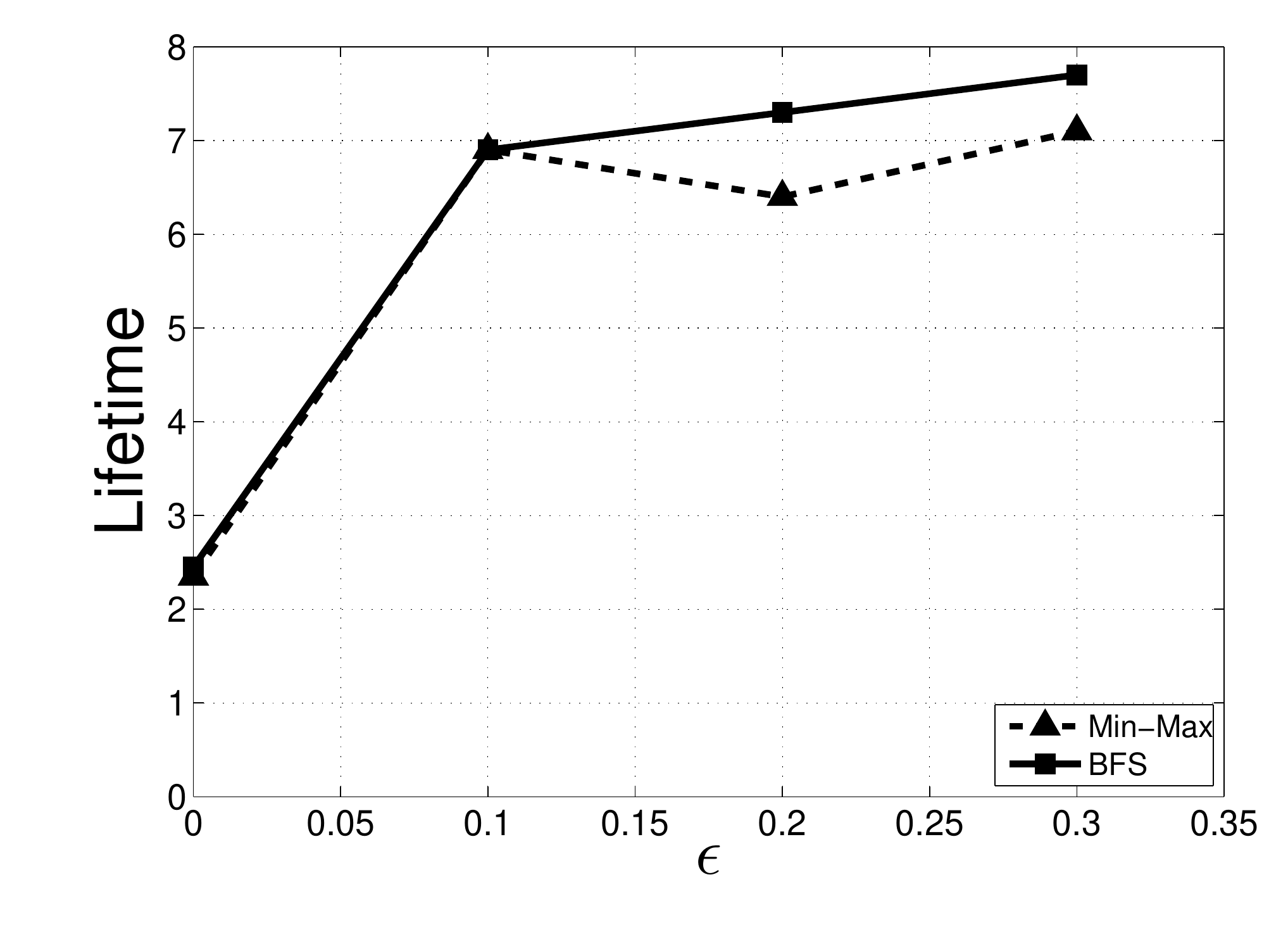}
  \label{k10i1ns}
  }
\caption{Lifetime vs $\epsilon$ when $\kappa=10$ and $Intensity=1$.}
\label{k10i1}
\end{minipage}
\hspace{0.5cm}
\begin{minipage}[b]{0.5\linewidth}
  \centering
  \subfloat[\sl{Uniform}]{
  \includegraphics[scale=0.35]{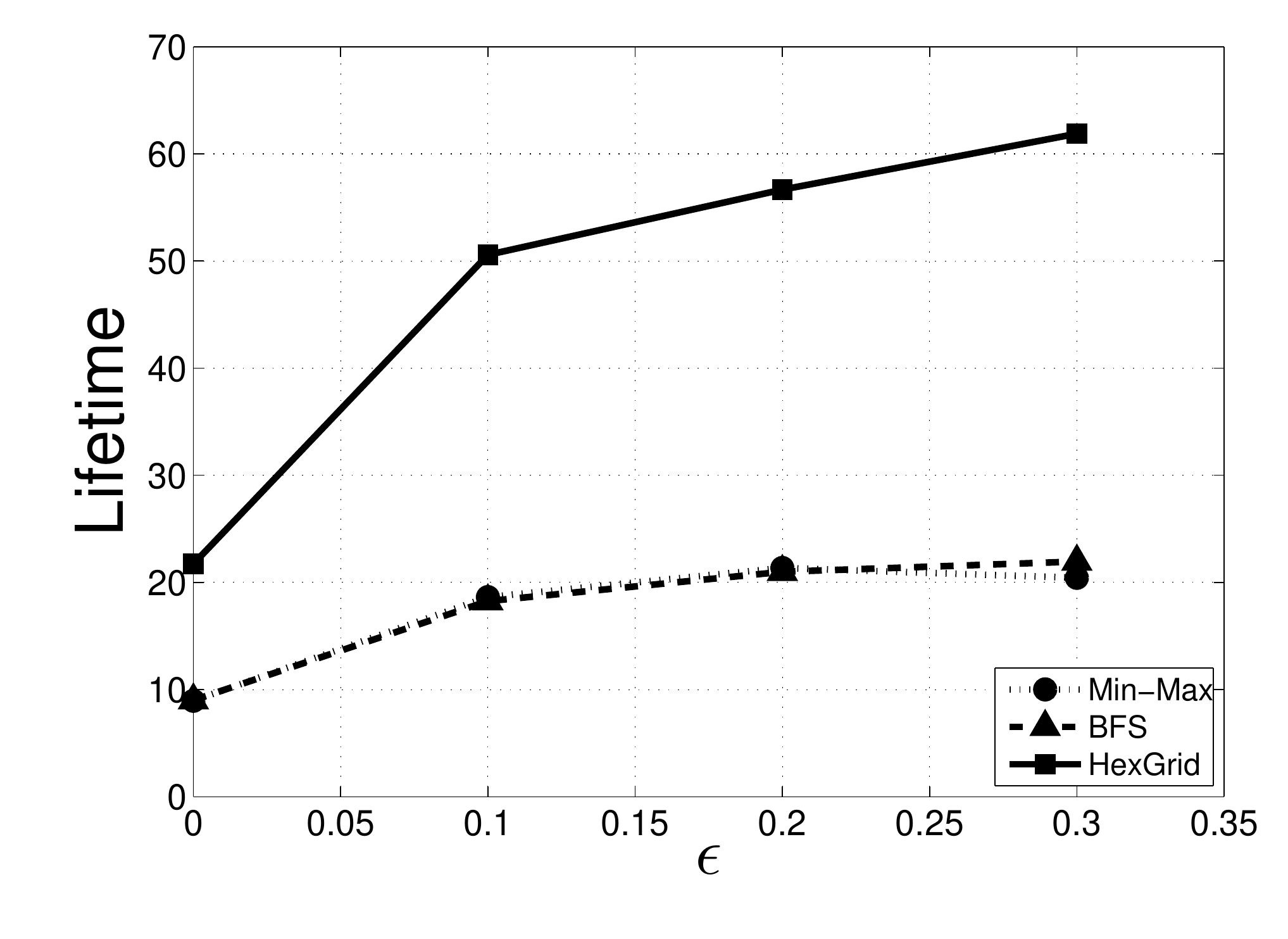}
  \label{k20i1un}
  }
  \hfil
  \subfloat[\sl{Non-Uniform}]{
  \includegraphics[scale=0.35]{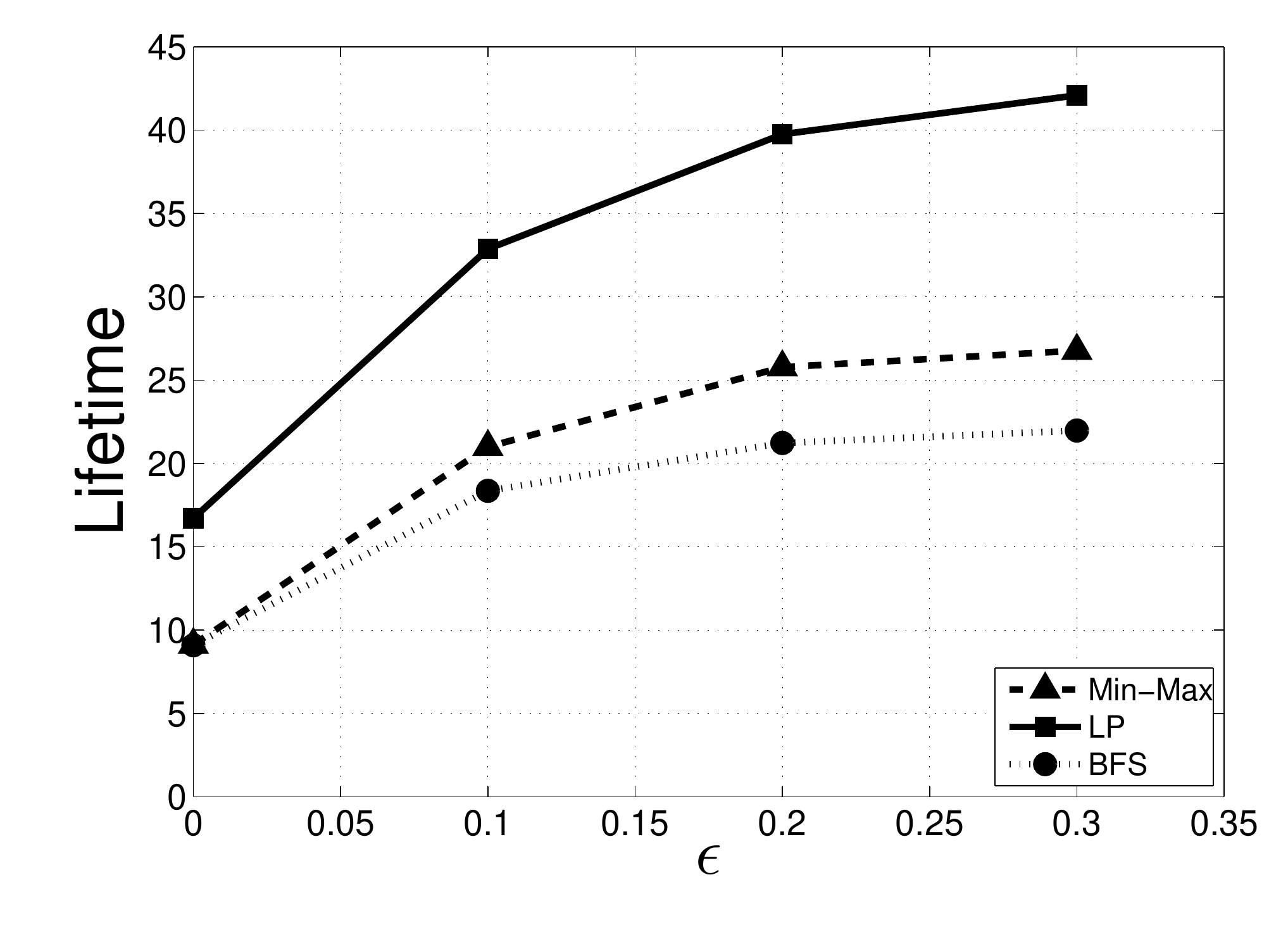}
  \label{k20i1de}
  }
  \hfil
  \subfloat[\sl{Non-Preemptive}]{
  \includegraphics[scale=0.35]{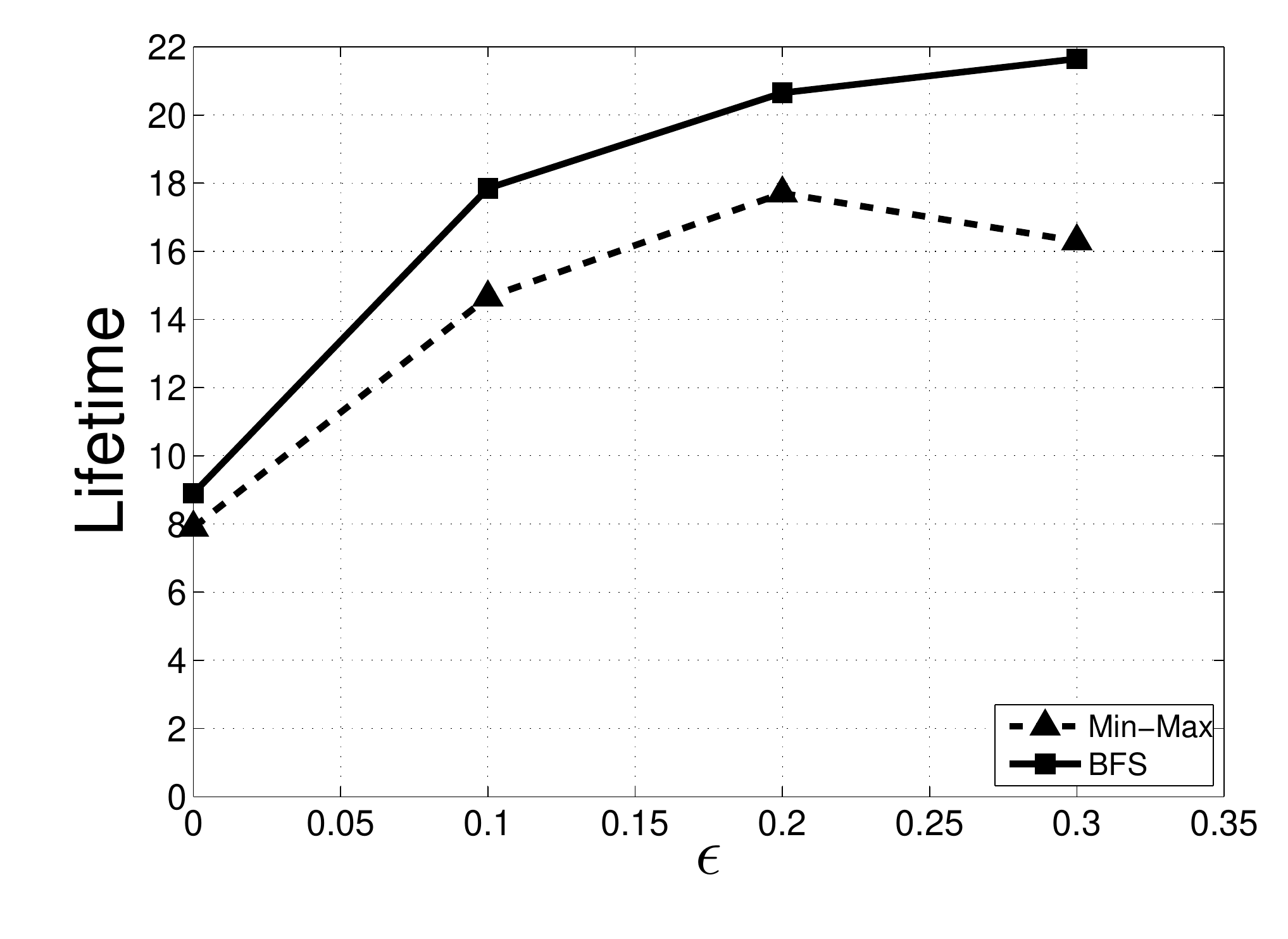}
  \label{k20i1ns}
  }
\caption{Lifetime vs $\epsilon$ when $\kappa=20$ and $Intensity=1$.}
\label{k20i1}
\end{minipage}
\end{figure}

\begin{figure}[p]
\begin{minipage}[b]{0.5\linewidth}
  \centering
  \subfloat[\sl{Uniform}]{
  \includegraphics[scale=0.35]{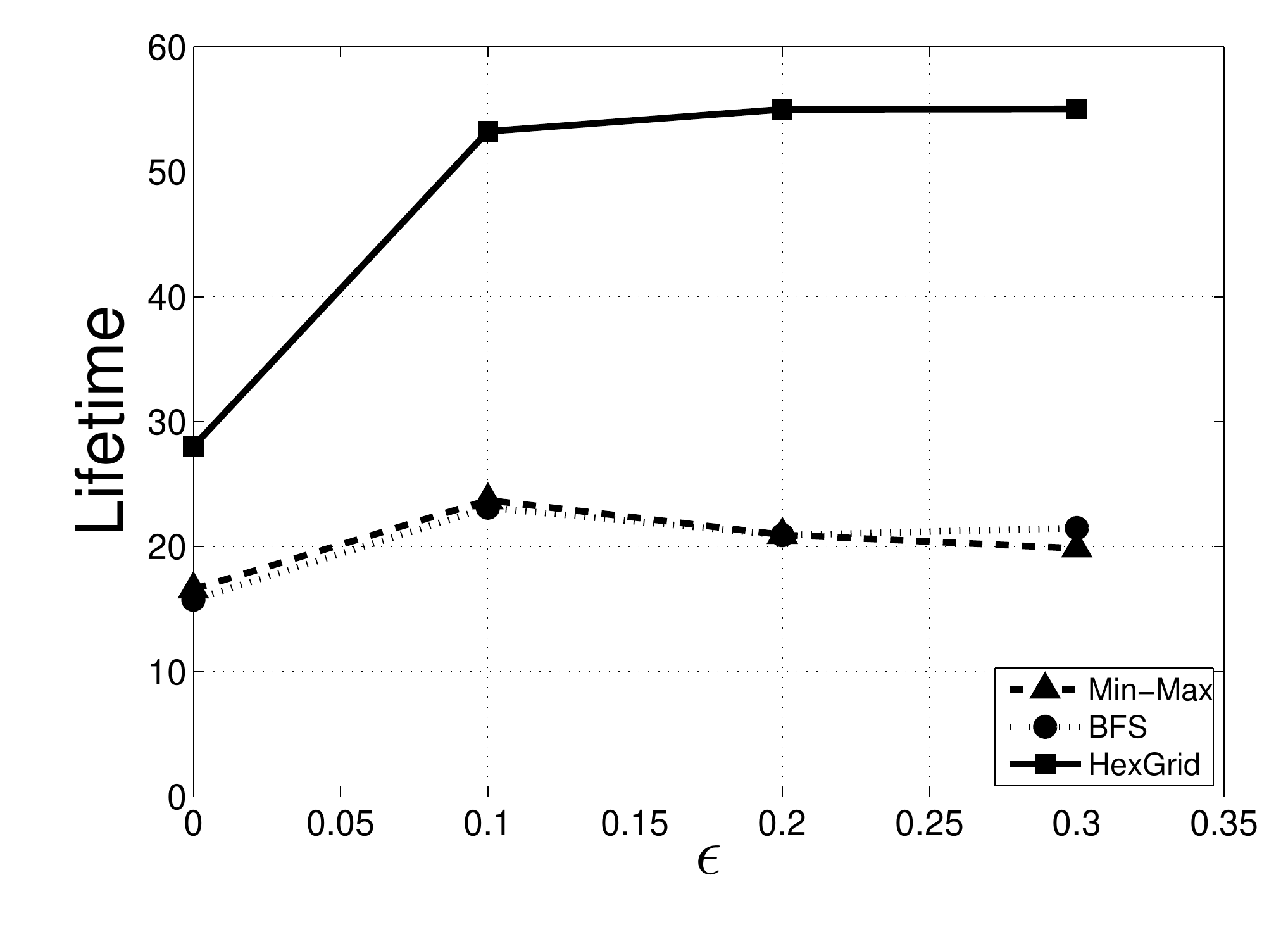}
  \label{k10i2un}
  }
  \hfil
  \subfloat[\sl{Non-Uniform}]{
  \includegraphics[scale=0.35]{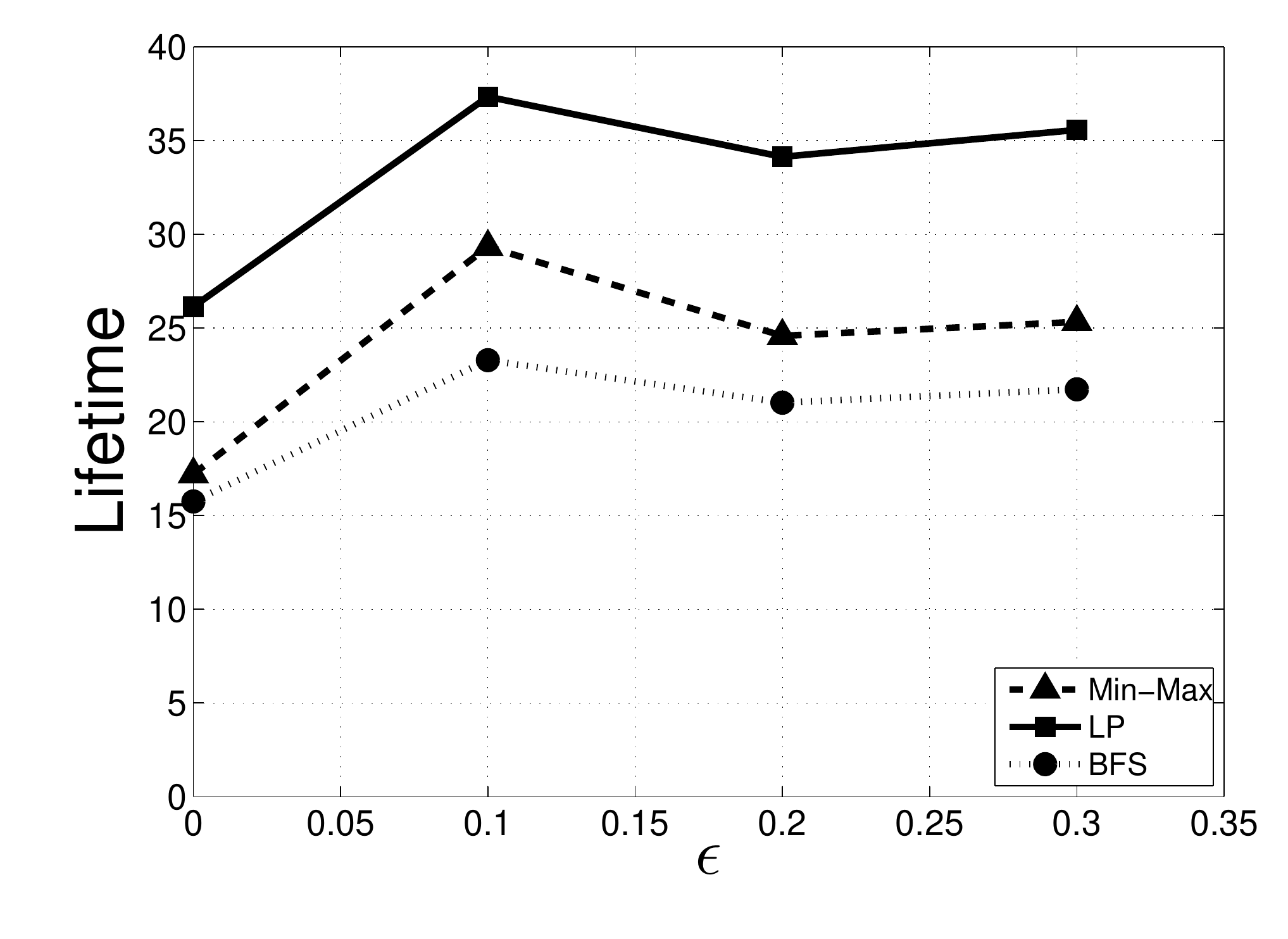}
  \label{k10i2de}
  }
  \hfil
  \subfloat[\sl{Non-Preemptive}]{
  \includegraphics[scale=0.35]{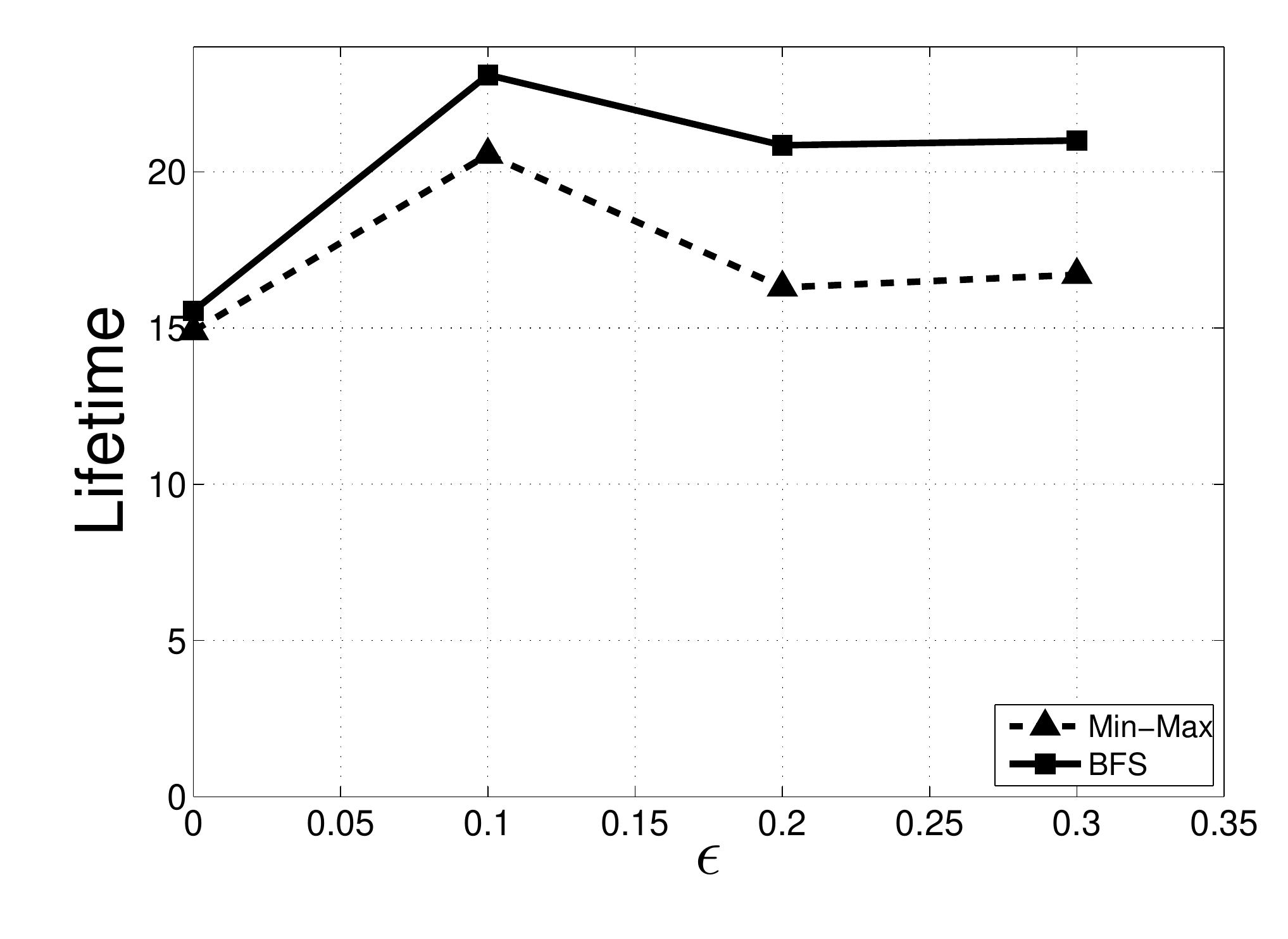}
  \label{k10i2ns}
  }
  \caption{Lifetime vs $\epsilon$ when $\kappa=10$ and $Intensity=2$.}
\label{k10i2}
\end{minipage}
\hspace{0.5cm}
\begin{minipage}[b]{0.5\linewidth}
  \centering
  \subfloat[\sl{Uniform}]{
  \includegraphics[scale=0.35]{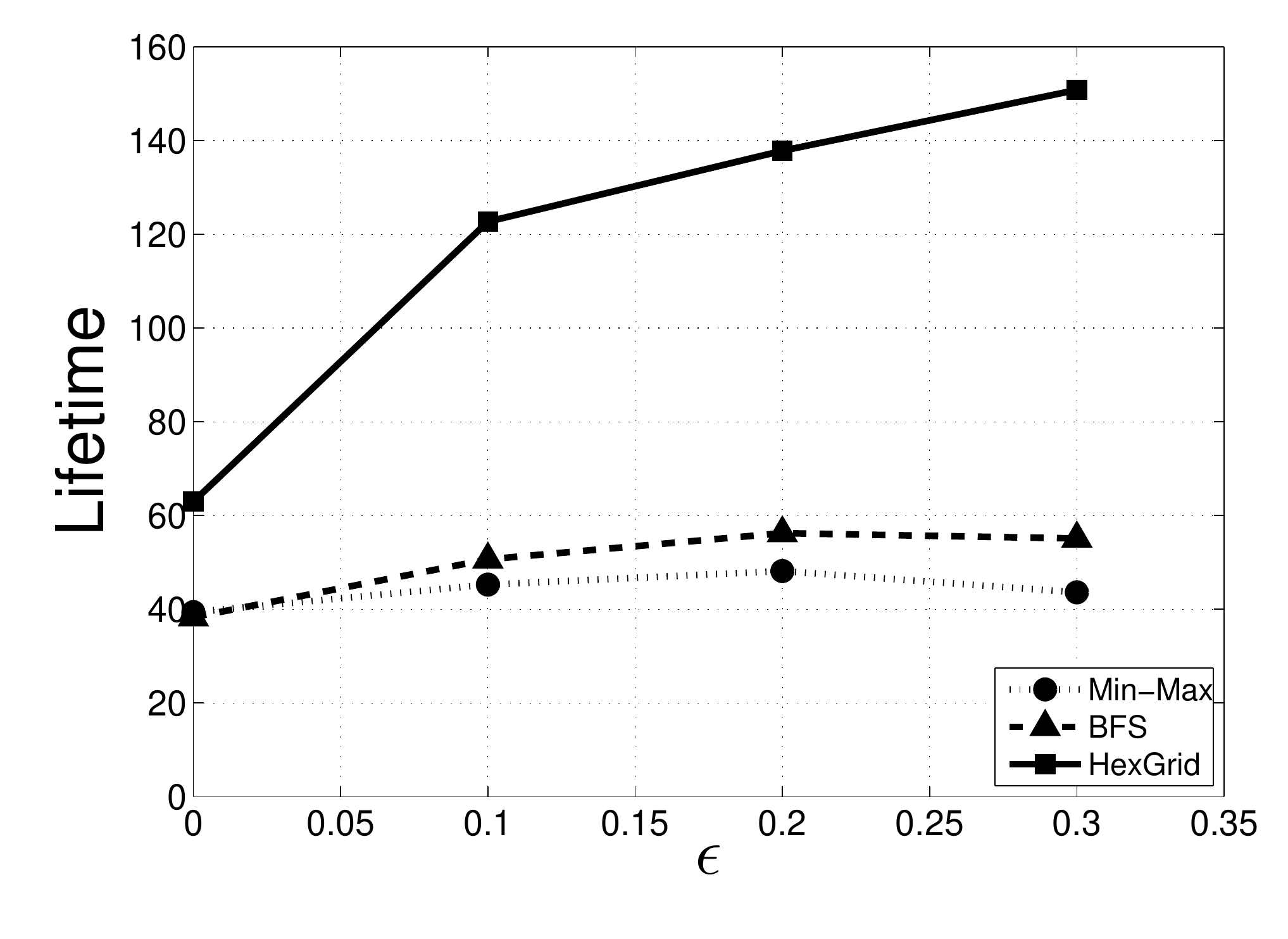}
  \label{k20i2un}
  }
  \hfil
  \subfloat[\sl{Non-Uniform}]{
  \includegraphics[scale=0.35]{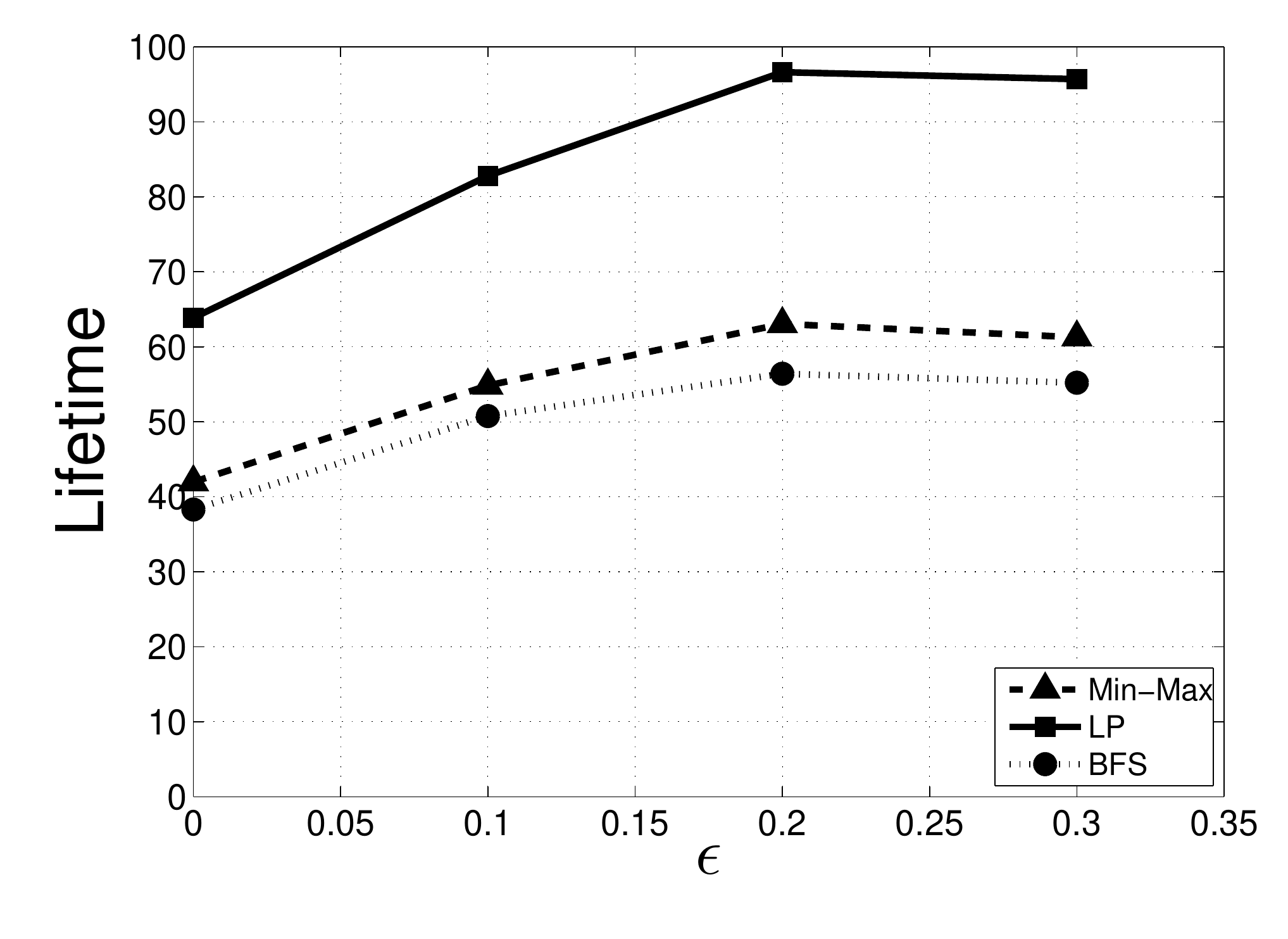}
  \label{k20i2de}
  }
  \hfil
  \subfloat[\sl{Non-Preemptive}]{
  \includegraphics[scale=0.35]{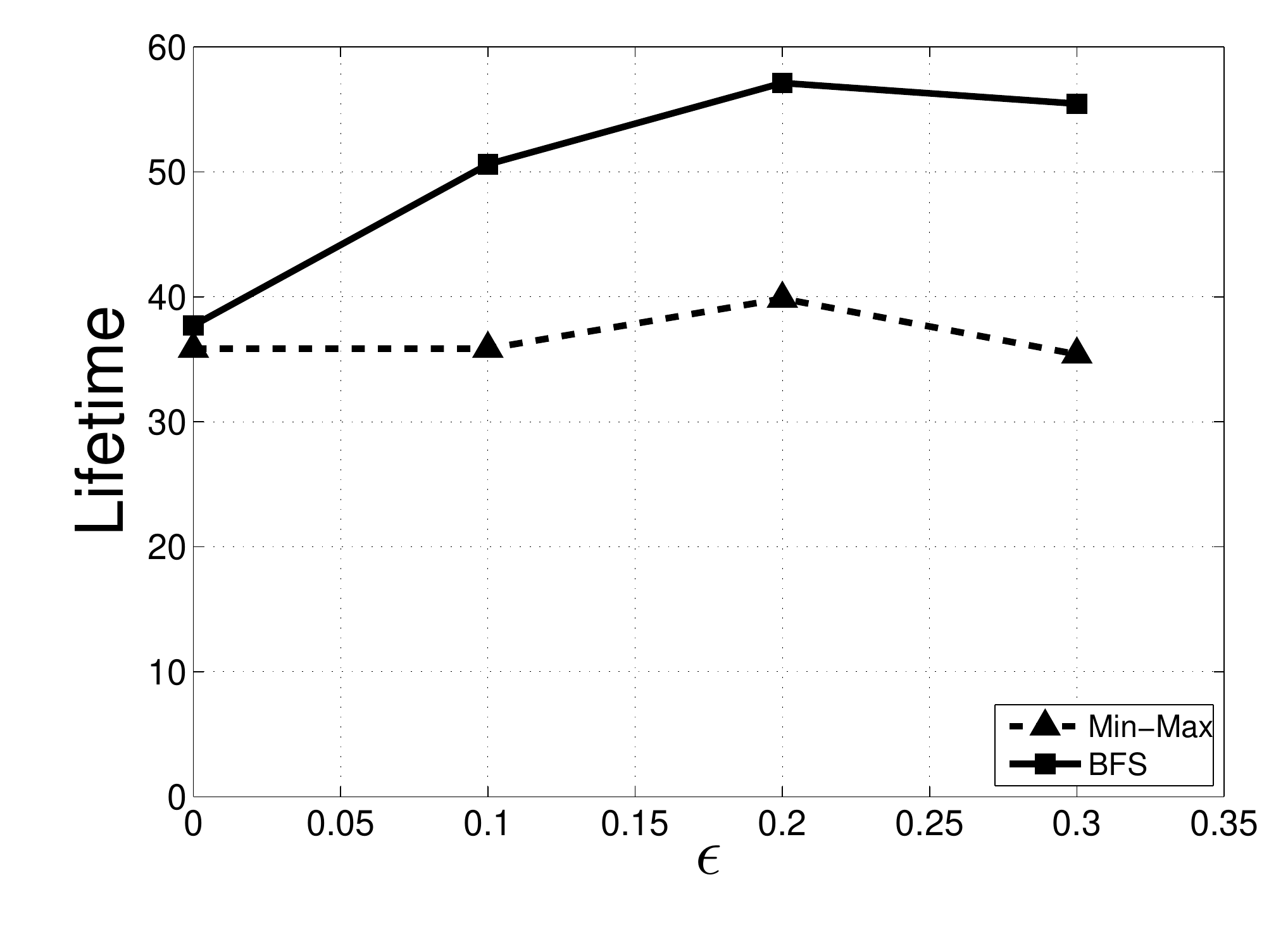}
  \label{k20i2ns}
  }
\caption{Lifetime vs $\epsilon$ when $\kappa=20$ and $Intensity=2$.}
\label{k20i2}
\end{minipage}
\end{figure}

In Figures \ref{k10i1}, \ref{k20i1}, \ref{k10i2} and \ref{k20i2}, \emph{Min-Max}, \emph{BFS} and \emph{LP} denote the corresponding algorithms executed on the square grid and \emph{HexGrid} denotes the BFS algorithm on the Hexagonal Grid. Figures \ref{k10i1} and \ref{k20i1} show the lifetime vs $\epsilon$ for $\kappa=10$ and $\kappa=20$ respectively when the intensity of the poisson process is $1$ and Figures \ref{k10i2} and \ref{k20i2} show the same for intensity $2$. In each of the figures, it may seen that the lifetime values increase sharply first followed by a decrease of slope in the curve. The net lifetime is, however, improved greatly over the lifetime when $\epsilon=0$, i.e., when the algorithms are exact. This decrease in slope may be attributed to the fact that, as $\epsilon$ increases from $\epsilon_{1}$ to $\epsilon_{2}$, the ratio of the number of sensors considered for $\epsilon_{1}$ to those considered for $\epsilon_{2}$ gets lower. This, in turn, seems to contribute directly to the decrease in slope.

Figures \ref{k10i1un}, \ref{k10i1un} and \ref{k10i1ns} and similarly, in Figures \ref{k10i1}, \ref{k20i1}, \ref{k10i2} and \ref{k20i2} show the distribution of lifetime vs $\epsilon$ in the case of the three activation policies. In all cases, the BFS seems to perform comparable to the Min-Max algorithm with better performance for the Non-Preemptive case. Hence, we are able to achieve similar performance with lower communication complexity. It may be seen in the Non-Uniform case (Figures \ref{k10i1de}, \ref{k20i1de}, \ref{k10i2de} and \ref{k20i2de}) that the LP algorithm performs much better than the others. This is in line with expectations since the LP provides close to optimal scheduling for covers found for a single grid, i.e.. Finally, in Figures \ref{k10i1un}, \ref{k20i1un}, \ref{k10i2un} and \ref{k10i2un}, we see that the hexagonal grid offers a very big jump in lifetime over the square grid algorithms, again, in line with the expectations considering the guarantees provided in Section \ref{sec:gridlifetimeguar}. 

\section{Conclusion and Discussion}
\label{sec:conc}

In this paper, we have investigated the problem of $\kappa$-weak coverage in sensor networks. The idea of using tilings to find $\kappa$-weak covers was introduced with lifetime guarantees provided for the Square and Hexagonal tilings. A novel LP algorithm was introduced to find covers for such tilings and its performance was compared against two other algorithms and was shown to achieve significantly better lifetime. The hexagonal grid was compared against the square and was also shown to achieve significantly better lifetime. Finally, an algorithm for finding covers in case GPS was not available was introduced.

\section{Acknowledgements}

The work in this paper was partially supported by NSF CAREER Grant 0348000.

\bibliographystyle{plain}
\bibliography{references}

\begin{thebibliography}{10}

\bibitem{ravi-coverage}
Ravi Balasubramanian, Srinivasan Ramasubramanian, and Alon Efrat.
\newblock Coverage time characteristics in sensor networks.
\newblock 2006.

\bibitem{sinha-trapcoverageinfocom09}
P.~Balister, Z.~Zheng, S.~Kumar, and P.~Sinha.
\newblock Trap coverage: Allowing coverage holes of bounded diameter in
  wireless sensor networks.
\newblock In {\em INFOCOM 2009, IEEE}, pages 136--144, April 2009.

\bibitem{fullcov-maxlife}
Jean Carle and David Simplot-Ryl.
\newblock Energy-efficient area monitoring for sensor networks.
\newblock {\em IEEE Computer}, 37:40--46, 2004.

\bibitem{ghosh-covsurvey}
Amitabha Ghosh and Sajal~K. Das.
\newblock Coverage and connectivity issues in wireless sensor networks: A
  survey.
\newblock {\em Pervasive and Mobile Computing}, 4(3):303 -- 334, 2008.

\bibitem{Ilyas-handbooksensornw}
Mohammad Ilyas and Imad Mahgoub.
\newblock {\em Handbook of Sensor Networks}.
\newblock CRC Press., 2004.

\bibitem{megerian-minexposurecov}
Seapahn Megerian, Farinaz Koushanfar, Gang Qu, Giacomino Veltri, and Miodrag
  Potkonjak.
\newblock Exposure in wireless sensor networks: theory and practical solutions.
\newblock {\em Wirel. Netw.}, 8(5):443--454, 2002.

\bibitem{meguer-cov}
Seapahn Meguerdichian, Farinaz Koushanfar, Gang Qu, and Miodrag Potkonjak.
\newblock Exposure in wireless ad-hoc sensor networks.
\newblock In {\em MobiCom '01: Proceedings of the 7th annual international
  conference on Mobile computing and networking}, pages 139--150, New York, NY,
  USA, 2001. ACM.

\bibitem{wan-kcov}
Peng-Jun Wan and Chih-Wei Yi.
\newblock Coverage by randomly deployed wireless sensor networks.
\newblock {\em IEEE/ACM Trans. Netw.}, 14(SI):2658--2669, 2006.

\bibitem{wu-covtime}
Jie Wu and Shuhui Yang.
\newblock Coverage issue in sensor networks with adjustable ranges.
\newblock {\em Parallel Processing Workshops, International Conference on},
  0:61--68, 2004.

\bibitem{xing-covsensornw}
Guoliang Xing, Xiaorui Wang, Yuanfang Zhang, Chenyang Lu, Robert Pless, and
  Christopher Gill.
\newblock Integrated coverage and connectivity configuration for energy
  conservation in sensor networks.
\newblock {\em ACM Trans. Sen. Netw.}, 1(1):36--72, 2005.

\bibitem{zhang-alphalifetime}
Honghai Zhang and Jennifer Hou.
\newblock On deriving the upper bound of {$\alpha$}-lifetime for large sensor
  networks.
\newblock In {\em MobiHoc '04: Proceedings of the 5th ACM international
  symposium on Mobile ad hoc networking and computing}, pages 121--132, New
  York, NY, USA, 2004. ACM.

\end{thebibliography}

\end{document}